\documentclass[showpacs,preprintnumbers,
superscriptaddress,amsmath,floatfix,prd]{revtex4}

\usepackage{amssymb}  
\usepackage{graphicx}

\usepackage[dvips]{color}

\usepackage[normalem]{ulem}  

\renewcommand\d{\delta}
\newcommand\e{\epsilon}

\newcommand{\non}{\nonumber\\}

\newcommand{\be}{\begin{equation}}
\newcommand{\ee}{\end{equation}}
\newcommand{\bea}{\begin{eqnarray}}
\newcommand{\eea}{\end{eqnarray}}
\newcommand{\ba}[1]{\begin{array}{#1}}
\newcommand{\ea}{\end{array}}

\begin{document}

\title{Critical temperature for kaon condensation in 
  color-flavor locked quark matter}

\author{Mark G.\ Alford}
\email{alford@wuphys.wustl.edu}
\affiliation{Department of Physics, Washington University St Louis, MO, 63130, USA}

\author{Matt Braby}
\email{mbraby@hbar.wustl.edu}
\affiliation{Department of Physics, Washington University St Louis, MO, 63130, USA}

\author{Andreas Schmitt}
\email{aschmitt@wuphys.wustl.edu}
\affiliation{Department of Physics, Washington University St Louis, MO, 63130, USA}

\date{November 21, 2007}

\begin{abstract}

We study the behavior of Goldstone bosons in color-flavor-locked (CFL)
quark matter at nonzero temperature. Chiral symmetry breaking in this
phase of cold and dense matter gives rise to pseudo-Goldstone bosons,
the lightest of these being
the charged and neutral kaons $K^+$ and $K^0$.  
At zero temperature, Bose-Einstein condensation of the kaons
occurs. Since all fermions are gapped, this kaon condensed CFL phase
can, for energies below the fermionic energy gap, be described
by an effective theory for the bosonic modes. We use this effective
theory to investigate the melting of the condensate: we determine the
temperature-dependent kaon masses self-consistently using the
two-particle irreducible effective action, and we compute the
transition temperature for Bose-Einstein condensation. Our results are
important for studies of transport properties of the kaon condensed
CFL phase, such as bulk viscosity.

\end{abstract}


\maketitle

\section{Introduction}
\label{intro}

Quark matter at sufficiently large densities and low temperatures is in the color-flavor-locked (CFL) state
\cite{Alford:1998mk}.
This state is a color superconductor \cite{bailin,reviews} since quarks form Cooper pairs, 
analogous to electrons in 
superconducting metals \cite{bcs}. 
The strong interaction provides an attractive channel in which condensation of quark Cooper pairs 
takes place. Many color-superconducting phases are thinkable, distinguished by their pairing pattern 
and physical properties. Only at asymptotically large densities do we know the ground state of three-flavor
quark matter with certainty. The reason is that at large densities all quark masses can be neglected
with respect to the quark chemical potential $\mu$. In this situation, the symmetry group of the
Lagrangian is $SU(3)_c\times SU(3)_R\times SU(3)_L \times U(1)_B$ with the color gauge group
$SU(3)_c$, the chiral symmetry group $SU(3)_R\times SU(3)_L$ and the baryon number
conservation group $U(1)_B$. Then, it is easy to see that the CFL phase is the only possible
color-superconducting phase in which all quasiquarks acquire an energy gap in their excitation spectrum.
Therefore, the CFL phase has the largest possible condensation energy, rendering this phase the ground state 
at asymptotic densities.

In the CFL phase, the symmetry group is spontaneously broken to the
residual global group $SU(3)_{c+R+L}$, generated by linear
combinations of the generators of the original groups. This residual
group locks color rotations with flavor rotations, hence the name
CFL. From the symmetry breaking pattern one concludes that there are
8+1 Goldstone bosons, an octet originating from the breakdown of the
chiral symmetry and a single mode originating from the breakdown of
$U(1)_B$. The latter is also termed ``superfluid mode'' since the
spontaneous breaking of the particle number conservation symmetry
accounts for the superfluidity of the CFL phase, just as in the case
of more conventional superfluids like $^4$He. For infinite density and
thus a ``full'' original chiral symmetry, these Goldstone bosons are
all massless. Therefore, the description of the physical properties of
the CFL phase, in particular the transport properties, is most
conveniently done within an effective theory approach
\cite{Casalbuoni:1999wu,Son:1999cm,Bedaque:2001je}. This approach is
valid for small energies where the excitation of any fermionic degrees
of freedom is suppressed due to the energy gap. Hence the CFL phase at
zero and sufficiently small temperatures can be, for most relevant
applications, thought of as a bosonic system of Goldstone modes. The
octet of Goldstone modes associated with chiral symmetry breaking has
the same quantum numbers as the corresponding pseudoscalar meson
octet in vacuum QCD.  Besides the effective field theory approach,
mesons in CFL have also been studied within Nambu-Jona-Lasinio models
\cite{Buballa:2004sx,Forbes:2004ww,Warringa:2006dk,Ebert:2006tc,Ruggieri:2007pi,Kleinhaus:2007ve}. 
In this approach, which models gluon exchange with a point-like
interaction, both fermionic quark modes and bosonic meson modes are
taken into account.

At moderate densities where the quark masses are no longer negligible,
the chiral symmetry is explicitly broken. This is certainly the case
for densities expected in the interior of neutron stars, where the
existence of deconfined quark matter is possible
\cite{perry,Alford:2006vz}.  For these densities, being on the order
of several times nuclear ground state density, we expect the quark
chemical potential $\mu$ to be of order $500$MeV.  The
strange quark mass is expected to be somewhere between the current
mass $\sim 100$MeV and the constituent quark mass $\sim 500$MeV and
can therefore not be neglected. Assuming a small explicit symmetry
breaking, the Goldstone bosons of the meson octet are still light,
however not massless. The order of the (zero
temperature) meson masses is inverted compared to the QCD vacuum
\cite{Son:1999cm}.  While the superfluid mode remains exactly
massless, the next lightest modes are expected to be the neutral and
charged kaons, $K^0$ and $K^+$. In the results of our study, we shall
focus exclusively on a system of $K^0$ and $K^+$, one of them being
condensed.

The situation in a compact star is further complicated by the
constraints of overall electric and color neutrality. Together with
the heaviness of the strange quark these constraints impose a stress
on the CFL phase and eventually lead to its breakdown. Whether this
breakdown occurs before the transition to hadronic matter is
unknown. If so, a more exotic color-superconducting phase has to take
over. Many possibilities for such a phase have been proposed, some of
them breaking rotational \cite{spin1} or translational \cite{LOFF}
symmetries. In particular, in the context of Goldstone boson
condensation, the possibility of a non-zero meson current has been
studied \cite{Schafer:2005ym}. Here we shall not discuss any of these
possibilities and solely consider CFL quark matter with an isotropic
kaon condensate. 


The calculations that we present in this paper are part of
an ongoing effort to
determine the ground state of quark matter at moderate densities 
using astrophysical observations. To this end, one has to compute the properties of candidate
color superconductors and test their compatibility with the data. 
Several approaches have been pursued in this direction, focussing on different transport properties
of different color superconductors. For instance, cooling properties of compact stars have been
related to the neutrino emissivity \cite{Carter:2000xf,Reddy:2002xc}, and glitches 
(sudden jumps in the rotation 
frequency of the star) have been connected to the rigidity of crystalline color-superconducting 
phases \cite{Mannarelli:2007bs}. Moreover, generic instabilities of so-called $r$-modes
\cite{Andersson:1997xt} require damping through sufficiently viscous fluids inside the stars in order
to be consistent with observed rotation frequencies. Therefore, shear viscosity \cite{Manuel:2004iv}
and bulk viscosity \cite{madsen} of unpaired and paired quark matter has recently caught attention.
Bulk viscosity has been studied for spin-1 color superconductors \cite{Sa'd:2006qv} as well as 
for color superconductors where strange quarks remain unpaired (2SC phase) \cite{Alford:2006gy}.

For the case of the CFL phase, a calculation has been performed where the kaon modes have been assumed
to be uncondensed \cite{Alford:2007rw} (for a calculation of the bulk viscosity in kaon condensed
nuclear matter, see Ref.\ \cite{Chatterjee:2007qs}). This is based on the assumption
of a sufficiently small kaon chemical potential which, given the uncertainties of this quantity, is
a possible but rather unlikely scenario. Therefore, it is important to study the opposite case where 
the kaon chemical potential is larger than the kaon mass, allowing for kaon condensation. It is then 
crucial to have a reliable estimate of the critical temperature for this condensation in order to 
perform calculations for instance of the bulk viscosity that are related to a certain temperature 
regime in the evolution of a compact star. In this paper, we develop a framework where 
both condensed and uncondensed kaons are described at nonzero temperatures and present 
a calculation of the transition temperature.

The paper is organized as follows. In Sec.\ \ref{seczeroT} we briefly
review the effective theory at zero temperature and discuss its
simplest extension to nonzero temperature, assuming an ideal gas of
mesons. This section mainly serves to introduce the notation and to
point out the difficulties that arise in the nonzero temperature
case. We introduce our self-consistent treatment of the
temperature-dependent masses in Sec.\ \ref{secphi4} for the case of a
simple $\varphi^4$ theory for a scalar field. This serves as the setup
for our analysis of mesons in CFL in Sec.\ \ref{seckaons}.  We derive
the self-consistent equations for the $K^0$ and $K^+$ masses and a
$K^0$ condensate, Sec.\ \ref{deriveq}, solve them numerically and present an
approximate analytical expression for the critical temperature. In Sec.\ \ref{secneutral} we
discuss the effect of electric neutrality and the presence of electrons on the critical temperature.

\section{Basic properties of mesons in CFL}
\label{seczeroT}

We start with explaining how mesons arise in the CFL phase and how they are described in 
an effective field theory. We briefly discuss the zero-temperature predictions of this theory
and present the simplest possible nonzero temperature approximation. 
Most of the contents of this section, which introduces the notation used in the subsequent sections, 
can be found in the existing literature. For more details about the following basic properties 
of mesons in CFL, see 
Refs.\ \cite{Casalbuoni:1999wu,Bedaque:2001je,Son:1999cm,Kaplan:2001qk,Schafer:2002ty}, and for more 
specific studies within the effective field theory see for instance Refs.\ \cite{Kryjevski:2004cw}.

\subsection{Meson fields from the CFL order parameter}

Diquark condensation takes place in the antisymmetric
antitriplet representation $[\bar{3}]_c$ of the color group $SU(3)_c$ (we neglect the much smaller
sextet gaps). Therefore, assuming condensation
in the antisymmetric spin singlet channel, we also have to choose the antisymmetric representation in flavor
space in order to ensure overall antisymmetry of the Cooper pair wave function. More precisely, 
the order parameter ${\cal M}$ lives in the antisymmetric antitriplet representation 
$[\bar{3}]_R\oplus[\bar{3}]_L$
of the chiral group $SU(3)_R\times SU(3)_L$,
\be
{\cal M} \in [\bar{3}]_c \otimes ([\bar{3}]_R\oplus[\bar{3}]_L) \, .
\ee
Let us denote the basis for the representations        
$[\bar{3}]_c$, $[\bar{3}]_R$, and $[\bar{3}]_L$ by $J_i$, $I_i^R$ and $I_i^L$, respectively ($i=1,2,3$),
where $(J_i)_{jk}=(I_i^R)_{jk}=(I_i^L)_{jk}=-i\e_{ijk}$.
Then, the gauge-variant 
order parameter for color superconductivity is given by a pair ${\cal M}=(\Delta^R,\Delta^L)$,
where $\Delta^R$, $\Delta^L$ are complex $3\times 3$ matrices,
\be \label{M2}
{\cal M} = \Delta_{ij}^R J_i\otimes I_j^R \oplus \Delta_{k\ell}^L J_k\otimes I_\ell^L \, .
\ee
In the CFL phase, the order parameter is given by
\be
\Delta^R_{ij}=\Delta^L_{ij}=\delta_{ij}
\ee
and consequently breaks the group
\be
G\equiv SU(3)_c\times SU(3)_R \times SU(3)_L
\ee
down to the residual group 
\be
H\equiv SU(3)_{c+R+L} \, .
\ee  
In other words, the ground state of the system is invariant under
transformations of the group $H$. It is therefore only invariant under
simultaneous left-handed flavor, right-handed flavor, and color
rotations.  Hence chiral symmetry is broken in the CFL phase. Usually,
i.e., in the hadronic phase of QCD, chiral symmetry breaking is
induced by a chiral condensate of the form
$\langle\bar{\psi}\psi\rangle$.  In the CFL phase chiral symmetry
breaking is induced by a diquark order parameter of the form
$\langle\psi\psi\rangle$. Here the breakdown of the chiral symmetries
occurs because of the locking to color degrees of freedom.

Other order parameters than the unit matrix 
$\Delta^R=\Delta^L={\bf 1}$ are possible. 
For instance, the so-called 2SC order parameter is
given by $\Delta^R_{ij}=\Delta^L_{ij}=\delta_{i3}\delta_{j3}$.  In
this case, left-handed and right-handed flavor 
groups are separately broken
down to $SU(2)_L$ and $SU(2)_R$, respectively. Thus the ground state
is still invariant under separate left-handed and right-handed
flavor rotations. We shall not elaborate on further possible structures of
the order parameter. For an exhaustive discussion of all possibilities
(with the restriction $\Delta^R=\Delta^L$), see
Ref.\ \cite{Rajagopal:2005dg}.

Without explicit symmetry breaking of $G$, the CFL ground state is
degenerate, i.e., it can be rotated by a transformation of the
degeneracy space $G/H$ into an equivalent one with the same physical
properties. A (small) explicit breaking of $G$ through nonzero quark
masses removes this degeneracy, and a rotation with an
element of $G/H$ may lead to a physically distinct state, a state 
where the Goldstone bosons form a Bose-Einstein condensate.
Rotations in $G/H$ are for example axial color or axial
flavor rotations. With $V\in SU(3)$, an axial flavor rotation on
${\cal M}=(\Delta^R,\Delta^L)$ is given by

\be
V({\cal M}) = (\Delta^RV,\Delta^L V^{-1}) \, ,
\ee
where $\Delta^RV$ and $\Delta^L V^{-1}$ are usual matrix multiplications. We see that a state 
with $\Delta^R=\Delta^L$ gets rotated into a state with $\Delta^R\neq\Delta^L$. 

One should mention that 
besides the chiral basis, also a basis of scalar and pseudoscalar order parameters is used in the 
literature \cite{Buballa:2004sx,Forbes:2004ww,Warringa:2006dk}. 
In this basis, the order parameter is a pair $(\Delta_s, \Delta_p)$,
where $\Delta^R=\Delta^L$ corresponds to a vanishing pseudoscalar component, $\Delta_p=0$. In other words,
for massless quarks one can, due to the degeneracy of the ground state, always choose $\Delta_p=0$.
Now, axial color or flavor transformations generate a nonzero $\Delta_p$. The explicit form of 
an axial flavor transformation is for instance $(\Delta_s,0)\to (\Delta_s\cos Q,i\Delta_s\sin Q)$
where $V=\exp(iQ)\in SU(3)$. For the following definition of the meson fields we shall return to the 
chiral basis and just remark that the definition (\ref{defsigma}) has the form 
$\Sigma=(\Delta_s-\Delta_p)^\dag(\Delta_s+\Delta_p)$ in the basis of scalar and pseudoscalar condensates.

The mesonic field is defined as 
\be \label{defsigma}
\Sigma\equiv (\Delta^L)^\dag \Delta^R \, .
\ee 
For unitary $\Delta^L$, $\Delta^R$, the $3\times 3$ matrix is unitary,
$\Sigma\in U(3)$. It has 9 degrees of freedom, one of which, the
$\eta'$, corresponds to breaking of the approximate $U(1)_A$
symmetry. Since we expect $U(1)_A$ to be explicitly broken at moderate
densities, we shall ignore the $\eta'$ so that $\Sigma\in SU(3)$.
Being a product of a left-handed anti-diquark (or hole-diquark)
condensate and a right-handed diquark condensate, $\Sigma$ carries
quantum numbers of four (anti-)colors and four (anti-)flavors. From
the above definition of the order parameter we conclude that for
example $\Delta_{12}^L$ is a Cooper pair condensate of two left-handed
quarks with colors 2,3 and flavors 1,3 (in an antisymmetric
combination), $\Delta_{13}^R$ is a Cooper pair condensate of two
right-handed quarks with colors 2,3 and flavors 1,2 etc. By carrying
out the matrix multiplication in Eq.\ (\ref{defsigma}) we find matrix
elements of the form $\Sigma_{12}=(\Delta_{11}^L)^*\Delta_{12}^R+
(\Delta_{21}^L)^*\Delta_{22}^R+(\Delta_{31}^L)^*\Delta_{32}^R$ etc.,
which can be translated into the respective quantum numbers. Ordering
colors as $(r,g,b)$ and flavors as $(u,d,s)$ (where we abbreviate red,
green, blue by $r$, $g$, $b$, and up, down, strange by $u$, $d$, $s$)
we obtain for example \bea \Sigma_{32} &\rightarrow&
\overline{gb,ud}_L \, gb,us_R + \overline{rb,ud}_L \, rb,us_R +
\overline{rg,ud}_L\, rg,us_R \rightarrow \overline{ud}_L\, us_R \, .
\eea We see that the colors cancel each other. This is expected from
the transformation properties of $\Sigma$. It transforms under the
color-flavor group $G$ as $\Sigma\to V_L^\dag\Sigma V_R$ where
$(V_R,V_L)\in SU(3)_R \times SU(3)_L$.  Hence $\Sigma$ is a color
singlet. For the whole matrix we find the flavor quantum numbers 
\be
\Sigma \rightarrow \left(\begin{array}{ccc} \overline{ds}_L\,ds_R \;\;
  & \overline{ds}_L\,us_R \;\;& \overline{ds}_L\,ud_R
  \\ \\ \overline{us}_L\,ds_R \;\;& \overline{us}_L\,us_R \;\;&
  \overline{us}_L\,ud_R \\ \\ \overline{ud}_L\,ds_R \;\;&
  \overline{ud}_L\,us_R \;\;& \overline{ud}_L\,ud_R
\end{array}\right) \, .
\label{4quark}
\ee
We see that the states are composed of four quarks of the structure
$\bar{q}\bar{q}qq$ as opposed to $\bar{q}q$ in the case of the usual
chiral condensate. Nevertheless, we find analogous quantum
numbers. For example, the charged kaon states correspond to 
$K^+=\bar{s}u$, $K^-=\bar{u}s$, 
generated by the Gell-Mann matrices
$T_4$, $T_5$, and the neutral kaon states to $K^0=\bar{s}d$,
$\bar{K}^0=\bar{d}s$, generated by $T_6$, $T_7$.

\subsection{Effective chiral Lagrangian and kaon condensation at zero temperature}

The effective Lagrangian for mesons in CFL is 
\cite{Son:1999cm,Bedaque:2001je,Kaplan:2001qk,Schafer:2002ty}
\bea \label{Leff}
{\cal L} &=& \frac{f_\pi^2}{4} {\rm Tr}\left[(\partial_0\Sigma+i[A,\Sigma])
(\partial_0\Sigma^\dag-i[A,\Sigma]^\dag) - v_\pi^2\partial_i\Sigma\partial_i\Sigma^\dag\right]
+\frac{af_\pi^2}{2}{\rm det}M{\rm Tr}[M^{-1}(\Sigma+\Sigma^\dag)] \, ,
\eea
where we have omitted electromagnetic corrections to the 
charged meson masses \cite{Kaplan:2001qk} and instanton effects \cite{Schafer:2002ty}.
The meson field, defined in Eq.\ (\ref{defsigma}) can be parametrized as  
\be
\Sigma = e^{i\theta/f_\pi} \, , 
\ee
where $\theta$ is a matrix in the Lie algebra of $SU(3)$.
We have defined the ``gauge field''  
\be \label{ALAR}
A\equiv \mu_Q Q -\frac{M^2}{2\mu} \, .
\ee
Here, $Q={\rm diag}(2/3,-1/3,-1/3)$ and $M={\rm diag}(m_u,m_d,m_s)$ are the electric charge and quark 
mass matrices in flavor space,
and $\mu_Q$ the chemical potential associated with electric charge.
At asymptotically high density
the constants $f_\pi$, $v_\pi$, and $a$ can be determined
by matching the effective theory to perturbative QCD \cite{Son:1999cm,Hong:1999ei},
\bea \label{matching}
f_\pi^2 &=& \frac{21-8\ln 2}{18}\frac{\mu^2}{\pi^2} \, , \qquad v_\pi = \frac{1}{\sqrt{3}} \, , \qquad
a=\frac{3\Delta^2}{\pi^2f_\pi^2} \, ,
\eea
where $\Delta$ is the fermionic energy gap at zero temperature. Since Lorentz invariance is broken in a 
medium, spatial and temporal components may have different prefactors, $v_\pi\neq 1$.

We shall subtract the CFL part of the Lagrangian, which is given by $\Sigma={\bf 1}$. Then, 
we arrive at the new Lagrangian (which we also denote by ${\cal L}$ since from now on this is the
only relevant Lagrangian)
\bea \label{L1}
{\cal L} &=& {\cal L}_0 + {\cal L}_1 \, ,
\eea
where 
\begin{subequations} \label{L0L1}
\bea
{\cal L}_0&\equiv& 
\frac{f_\pi^2}{2} {\rm Tr}\left[A^2-\left(A\cos \theta'\right)^2-\left(A\sin \theta'\right)^2 -2a
({\rm det}M)M^{-1}\left(1-\cos \theta'\right)\right] \, , \\
{\cal L}_1 &\equiv& \frac{f_\pi^2}{4}{\rm Tr}\left[\left(\partial_0\cos \theta'\right)^2 + 
\left(\partial_0\sin \theta'\right)^2
-v_\pi^2\left[\left(\nabla\cos \theta'\right)^2+
\left(\nabla\sin \theta'\right)^2\right]\right] \non
&&+\;i\,\frac{f_\pi^2}{2}{\rm Tr}\left[\left(\partial_0\cos \theta'\right)\left[A,\cos \theta'\right]
+\left(\partial_0\sin \theta'\right)\left[A,\sin \theta'\right]\right]
\, ,
\eea
\end{subequations}
with the dimensionless matrix field $\theta'\equiv\theta/f_\pi$.
Now we separate the zero-momentum mode of the field, $\theta(K)\to \theta(0) + \theta(K)$
where $K=(k_0,{\bf k})$ is the four-momentum with the bosonic Matsubara frequencies
$\omega_n=ik_0=2n\pi T$. This separation is necessary to take into account Bose 
condensation as usual (= as in the case of a simple one-component ideal Bose gas \cite{kapusta}).
Then, one approximates the zero mode by its expectation value
\be
\theta(0) = \langle\theta(0)\rangle \equiv \Phi \equiv \phi_a T_a \, ,
\ee  
where $T_a$, ($a=1,\ldots ,8$) are the Gell-Mann matrices.
In this subsection, we only discuss the zero-temperature case for which we only need ${\cal L}_0$ with 
$\theta$ replaced by the condensate $\Phi$.
In the subsequent sections we shall then turn to the remaining terms of ${\cal L}_0$ that include the
fluctuations $\theta$ and the derivative terms in ${\cal L}_1$.

\begin{figure} [ht]
\begin{center}
\includegraphics[width=0.5\textwidth]{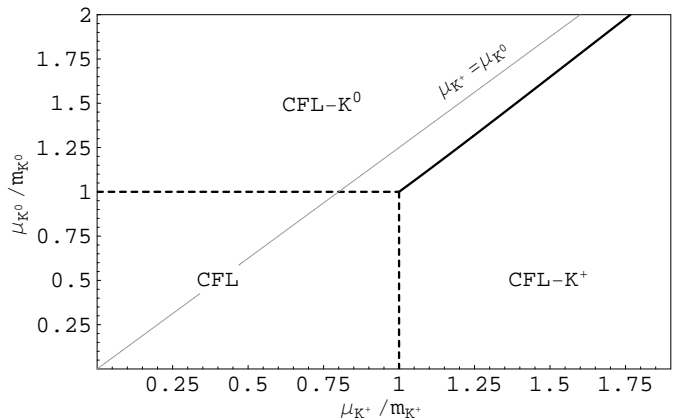}
\caption{
Zero-temperature phase diagram for kaon condensation in the $\mu_{K^+}$-$\mu_{K^0}$-plane. 
No condensation occurs if the chemical potential is smaller than the meson mass. Coexistence of the two 
condensates is only possible along the (solid) line that separates the CFL-$K^0$ from the CFL-$K^+$ phase. 
This line is given by Eq.\ (\ref{curve}) and marks a first order phase transition. For large chemical 
potentials, it approaches the line $\mu_{K^+}=\mu_{K^0}$. The (dashed) lines separating 
either of 
the two meson condensed phases from the pure CFL phase are second order phase transition lines. In the 
condensed phases, the condensate and free energy are given by Eqs.\ (\ref{costheta}) and 
(\ref{freeenergy}), respectively.} 
\label{figmu1mu20}
\end{center}
\end{figure}

For simplicity, we consider only kaons, i.e., $a=4,5,6,7$. Physically, this is motivated by the 
fact that we expect the kaons to be the lightest mesonic degrees of freedom. Moreover, 
we fix the phase of the condensates by setting $\phi_5=\phi_7=0$. Then, we denote $\phi_{K^+}\equiv
\phi_4$, $\phi_{K^0}\equiv\phi_6$, and
\be
\phi\equiv \sqrt{\phi_{K^+}^2+\phi_{K^0}^2} \, .
\ee
In this simple case, the matrix $\Phi$ is has the property $\Phi^3=\phi^2\,\Phi$.
Consequently, 
\bea
\cos \frac{\Phi}{f_\pi} = 1-\left(1-\cos\frac{\phi}{f_\pi}\right)\frac{\Phi^2}{\phi^2} \, , \qquad
\sin \frac{\Phi}{f_\pi} = \frac{\Phi}{\phi}\sin\frac{\phi}{f_\pi} \, .
\eea
We may use these identities to compute the zero-temperature free energy density $U$, which is simply the
negative of the constant terms in the Lagrangian. After performing
the trace in flavor space, we obtain 
\bea \label{Uallorder}
U &=& f_\pi^2\left(1-\cos\frac{\phi}{f_\pi}\right)\left[(m^2_{K^+}-\mu^2_{K^+})
\frac{\phi^2_{K^+}}{\phi^2} + (m^2_{K^0}-\mu^2_{K^0})
\frac{\phi^2_{K^0}}{\phi^2}\right] \non
&&+\;\frac{f_\pi^2}{2}\left(1-\cos\frac{\phi}{f_\pi}\right)^2
\left(\mu_{K^+}\frac{\phi^2_{K^+}}{\phi^2}+\mu_{K^0}
\frac{\phi^2_{K^0}}{\phi^2}\right)^2  \, , 
\eea
where we have defined the kaon chemical 
potentials and masses (squared)
\begin{subequations} \label{masschem}
\bea
\mu_{K^+}&\equiv&\mu_Q+\frac{m_s^2-m_u^2}{2\mu} \, , \qquad \mu_{K^0} \equiv \frac{m_s^2-m_d^2}{2\mu} \, , \\
m_{K^+}^2&\equiv& am_d(m_s+m_u) \, , \qquad m_{K^0}^2 \equiv  am_u(m_s+m_d)
\, .
\eea
\end{subequations}
We see that these quantities simply arise from evaluating the Lagrangian (\ref{Leff}), which, in turn, 
can be derived from symmetry arguments. (Calculations of the meson masses and chemical potentials 
using an NJL model are in accordance with these expressions \cite{Kleinhaus:2007ve}.) 
The free energy can be used to find the ground state of the system for arbitrary chemical 
potentials $\mu_{K^+}$, $\mu_{K^0}$. To this end, one has to minimize the free energy through the equations
\be \label{saddle}
\frac{\partial U}{\partial\phi_{K^+}}=\frac{\partial U}{\partial\phi_{K^0}}=0 \, .
\ee
By construction, the free energy of the CFL state is given by $U=0$, corresponding to 
$\phi_{K^+}=\phi_{K^0}=0$. If one of the condensates vanishes, one of the equations (\ref{saddle}) is 
automatically fulfilled. The solution of the other one is
\be \label{costheta}
\cos\frac{\phi_i}{f_\pi} = \left\{\begin{array}{cc} 1 & {\rm for}\; m_i^2>\mu_i^2 \\ \frac{m_i^2}{\mu_i^2} 
& {\rm for}\; m_i^2<\mu_i^2 \end{array}\right. \, , \qquad i=K^+,K^0 \, ,
\ee
and the free energy density is
\be \label{freeenergy}
U= \left\{\begin{array}{cc} 0 & {\rm for}\; m_i^2>\mu_i^2 \\ 
-\frac{f_\pi^2(m_i^2-\mu_i^2)^2}{2\mu_i^2} & {\rm for}\; m_i^2<\mu_i^2 \end{array}\right. \, , 
\qquad i=K^+,K^0 \, .
\ee
By equating the free energies of the two phases $\phi_{K^+}=0$, $\phi_{K^0}\neq 0$ and 
$\phi_{K^+}\neq 0$, $\phi_{K^0}= 0$
one finds the condition for coexistence of two condensates,
\be \label{curve}
\mu_{K^0}^2(\mu_{K^+}^2 - m_{K^+}^2)^2 =\mu_{K^+}^2(\mu_{K^0}^2 - m_{K^0}^2)^2 \, .
\ee
This condition can also be obtained by assuming two nonvanishing condensates in Eqs.\ (\ref{saddle}).
This results in a phase diagram shown in Fig.\ \ref{figmu1mu20}, where we restrict ourselves to 
$\mu_{K^+},\mu_{K^0}>0$ without loss of generality. Note that we have not included any 
neutrality constraint, i.e., all states in the phase diagram where there is a charged kaon condensate are 
not neutral. We shall use the phase diagram in Fig.\ \ref{figmu1mu20} as a basis for our nonzero temperature
results in Sec.\ \ref{seckaons}. 

\subsection{Expansion in the meson fields and ideal gas approximation}

Before turning to the self-consistent treatment of nonzero temperature effects, let us briefly 
discuss the expansion of the Lagrangian in the fields. We shall collect terms up to fourth 
order in the field. In this subsection, only the second order terms shall be evaluated. 
We shall see that in this simple approximation the system is described by decoupled meson 
components, each corresponding to an ideal Bose gas. Then in Sec.\ \ref{seckaons} we include the fourth-order
terms into our analysis.
  
Up to fourth order in the matrix-valued field $\theta$, we obtain from the Lagrangian, given in 
Eqs.\ (\ref{L1}) and (\ref{L0L1}), 
\begin{subequations}
\bea
{\cal L}_0 &=& \frac{1}{2}{\rm Tr}[X\theta^2-(A\theta)^2] 
+\frac{1}{2f_\pi^2}{\rm Tr}\left[\frac{1}{3}(A\theta)^2\theta^2-\frac{1}{12}X\theta^4
-\frac{1}{4}(A\theta^2)^2\right] \, ,\\
{\cal L}_1 &=& \frac{1}{4}{\rm Tr}\Big[(\partial_0\theta)^2-v_\pi^2
(\nabla \theta)^2+2i(\partial_0\theta)[A,\theta]\Big] \, , 
\eea
\end{subequations} 
where we abbreviated 
\be
X\equiv A^2-a({\rm det}M)M^{-1} \, .
\ee
We have neglected all derivative terms of order $\theta^4$ such as 
${\rm Tr}[(\theta\partial_0\theta)^2]$ etc. 
After introducing the condensate by shifting the fields as explained above, $\theta\to\theta+\Phi$,
the Lagrangian becomes 
\be
{\cal L} = -U(\Phi)+{\cal L}^{(2)} + {\cal L}^{(3)} +{\cal L}^{(4)} \, .
\ee
Here the tree-level potential for the condensate is 
\bea \label{treeU}
U(\Phi) &=& \frac{1}{2}{\rm Tr}[X\Phi^2-(A\Phi)^2]
+\frac{1}{2f_\pi^2}{\rm Tr}
\left[\frac{1}{3}(A\Phi)^2\Phi^2-\frac{1}{12}X\Phi^4-\frac{1}{4}(A\Phi^2)^2\right] \, .
\eea 
The terms quadratic in $\theta$ are 
\be
{\cal L}^{(2)} = {\cal L}^{(2)}_0+{\cal L}^{(2)}_1 \, , 
\ee
with non-derivative and derivative contributions
\begin{subequations}\label{quad}
\bea
{\cal L}^{(2)}_0 &\equiv& \frac{1}{2}{\rm Tr}[X\theta^2-(A\theta)^2]+\frac{1}{2f_\pi^2}
{\rm Tr}\Big[\frac{1}{3}[(A\theta)^2\Phi^2
+\{A\theta,A\Phi\}\{\theta,\Phi\}+(A\Phi)^2\theta^2]\non
&&-\;\frac{1}{12}X[\{\theta^2,\Phi^2\}+\{\theta,\Phi\}^2]
-\frac{1}{4}[2A\theta^2A\Phi^2
+(A\{\theta,\Phi\})^2]\Big] \, ,  \label{L02}\\
{\cal L}^{(2)}_1 &\equiv& \frac{1}{4}{\rm Tr}\Big[(\partial_0\theta)^2-v^2_\pi(\nabla \theta)^2 
+2i(\partial_0\theta)[A,\theta]\Big] \, .\label{L12}
\eea
\end{subequations}
Here, $\{-,-\}$ denotes the anticommutator. 
We also find induced cubic interactions,
\bea \label{cube}
{\cal L}^{(3)} &=& \frac{1}{2f_\pi^2}{\rm Tr}\Big[\frac{1}{3}[
\{A\theta,A\Phi\}\theta^2+(A\theta)^2\{\theta,\Phi\}]
-\frac{1}{12}X\{\theta^2,\{\theta,\Phi\}\}-\frac{1}{2} A\theta^2A\{\theta,\Phi\}\Big] \, , 
\eea
and finally quartic terms
\be
{\cal L}^{(4)} = \frac{1}{2f_\pi^2}{\rm Tr}\Big[\frac{1}{3}(A\theta)^2\theta^2-
\frac{1}{12}X\theta^4-\frac{1}{4}(A\theta^2)^2\Big] \, .  \label{L21} 
\ee
First, we only consider terms quadratic in the fields, i.e., terms $\propto \Phi^2, \theta^2$.
This corresponds to the first trace in $U(\Phi)$, Eq.\ (\ref{treeU}), the first trace in 
${\cal L}_0^{(2)}$, Eq.\ (\ref{L02}), and ${\cal L}_1^{(2)}$, Eq.\ (\ref{L12}).    
We obtain
\bea
U(\Phi) &=& \frac{1}{2}\Big[(m_{\pi^+}^2-\mu_{\pi^+}^2)(\phi_1^2+\phi_2^2)+
(m_{K^+}^2-\mu_{K^+}^2)(\phi_4^2+\phi_5^2)+
(m_{K^0}^2-\mu_{K^0}^2)(\phi_6^2+\phi_7^2) \non
&&+\; \frac{4am_um_d}{3}\phi_8^2+
am_um_s\left(\phi_3-\frac{\phi_8}{\sqrt{3}}\right)^2+
am_dm_s\left(\phi_3+\frac{\phi_8}{\sqrt{3}}\right)^2\Big] \, ,
\eea
where $\mu_{\pi^+}\equiv\mu_Q+(m_d^2-m_u^2)/(2\mu)$, $m_{\pi^+}^2\equiv am_s(m_d+m_u)$.
From the terms quadratic in $\theta$ we compute the free inverse propagator $S_0^{-1}$, which is an  
$8\times 8$ block diagonal matrix. The charged pion, 
charged kaon, and neutral kaon $2\times 2$ blocks read 
\bea \label{Sideal}
[S_0^{-1}]_{ab}=\left(\begin{array}{cc} -k_0^2+v_\pi^2 k^2+
m_i^2 -\mu_i^2 & -2ik_0\mu_i \\[1ex] 2ik_0\mu_i & -k_0^2+v_\pi^2 k^2+
m_i^2 -\mu_i^2 \end{array}\right) \, , \qquad i=\left\{\begin{array}{cc} \pi^+ & \mbox{for} \; a,b=1,2
\\ K^+ & \mbox{for} \; a,b=4,5\\ K^0 & \mbox{for} \; a,b=6,7 \end{array}\right. \, .
\eea
The remaining $2\times 2$ block 
corresponding to $\pi^0$, $\eta$ ($a,b=3,8$) is
\be
[S_0^{-1}]_{ab}=\left(\begin{array}{cc} -k_0^2+v_\pi^2 k^2+
am_s(m_u+m_d)  & 2 a m_s(m_d-m_u)/\sqrt{3} \\[1ex] 2 a m_s(m_d-m_u)/\sqrt{3} & -k_0^2+v_\pi^2 k^2+
a m_s(m_u+m_d)/3+4am_um_d/3 \end{array}\right) \, .
\ee 
Let us focus on the kaon part of the action. The inverse propagator in Eq.\ (\ref{Sideal}) is obviously the 
inverse free propagator of an ideal Bose gas with chemical potential $\mu_i$. 
Consequently, we can immediately deduce
the thermodynamic potential density $\Omega = -T\ln Z$, where $Z$ is the partition function. We obtain
\bea \label{Omega0}
\Omega &=& \frac{1}{2}(m_{K^+}^2-\mu_{K^+}^2)\phi_{K^+}^2+\frac{1}{2}(m_{K^0}^2-\mu_{K^0}^2)\phi_{K^0}^2 
+T\sum_{e=\pm}\sum_{i=K^+,K^0}\int\frac{d^3{\bf k}}{(2\pi)^3}\ln\left(1-e^{-\epsilon_i^e/T}\right) \, , 
\eea
where 
\be \label{Eie}
\epsilon_i^e\equiv \sqrt{v_\pi^2k^2+m_i^2}-e\,\mu_i \, , \qquad i=K^+,K^0 \, .
\ee
Here, $e=\pm$ accounts for particle and antiparticle degrees of freedom. In other words, $K^+$ 
has a chemical potential $+\mu_{K^+}$ whereas $K^-$ has a chemical 
potential $-\mu_{K^+}$ (and analogous for the particle/antiparticle pair $K^0$, $\overline{K}^0$). 
Note that the zero-temperature part of $\Omega$ can also be obtained from expanding 
the all-order expression in Eq.\ (\ref{Uallorder}). 

Let us comment on the validity of the ideal gas (free particle)
approximation. In
Fig.\ \ref{figmu1mu20} we saw that condensation at $T=0$ takes place
for $\mu_i^2>m_i^2$. The ideal gas approximation, however, restricts
the chemical potentials to values $\mu_i^2\leqslant m_i^2$. 
If the chemical potential of any boson becomes larger than its mass
then the low-momentum modes have
unphysical negative energies $\epsilon_i^+$. Therefore, one cannot use the
ideal gas approximation for these chemical potentials. Note that,
although the chemical potential is restricted, the ideal gas can
account for arbitrarily large particle number densities by increasing
the density in the condensate at $\mu_i=m_i$. In the physical context
of interest, however, the mesonic chemical potentials and
zero-temperature masses are given and may very well imply
$\mu_i^2>m_i^2$.  This means that 
we cannot use the ideal gas approximation
to describe both condensation and nonzero temperature effects.  The
simplest extension of this approximation also fails: suppose we also
take into account the second trace in Eq.\ (\ref{L02}). This leads to
corrections to the free propagator $S_0^{-1}$.  These corrections are
proportional to the condensates $\phi_i^2$. For small temperatures
then the above problem is cured: the resulting bosonic dispersions
$\epsilon_i^e$ will be positive for all three-momenta ${\bf k}$. However,
increasing the temperature leads to a decrease of the condensate.
Consequently, the dispersions approach the ideal gas dispersions for
$T\to T_c$ and at sufficiently large temperatures (and still $T<T_c$)
negative occupation numbers will occur whenever we consider the case
$\mu_i>m_i$. 
The fundamental problem with these approximations is that they
are not self-consistent: the excitations of the field
around its ground state expectation value
do not feel the same potential as the expectation value itself
feels. To fix this problem we must set up a
more elaborate approximation scheme which self-consistently determines
the nonzero-temperature meson masses, following well-known methods in
the context of general bosonic systems \cite{kapusta,Dolan:1973qd}.

\section{Self-consistent boson masses in $\varphi^4$ theory}
\label{secphi4}

Before we come back to mesons in CFL, we introduce our method with the simple example of 
$\varphi^4$ theory for a scalar field \cite{kapusta,Dolan:1973qd}. We shall employ the 
Cornwall-Jackiw-Tomboulis (CJT) formalism \cite{2PI}, as done for example in a similar context
of a linear sigma model \cite{Roder:2003uz}. In contrast to Ref.\ \cite{Roder:2003uz} we have to 
deal with a fixed chemical potential which leads to slightly different self-consistency equations.
For a discussion of the linear sigma model with chemical potential at zero temperature
see for instance Ref.\ \cite{Brauner:2006xm}. Nonvanishing chemical potentials in the CJT formalism 
have been considered at nonzero temperature in Ref.\ \cite{Andersen:2006ys}, where a $1/N$ expansion 
in the number of fields $N$ is employed. 
We discuss the simple case of $\varphi^4$ theory in detail because we shall reduce the chiral 
effective theory for mesons in CFL essentially to a $\varphi^4$ theory, so that we can directly 
apply the results of this section to the meson case.

\subsection{Lagrangian and 2PI effective action}

We start from the Lagrangian for a complex scalar field $\varphi$ with chemical potential $\mu$, mass $m$,
and coupling constant $\lambda$,
\be
{\cal L} = |(\partial_0-i\mu)\varphi|^2-|\nabla\varphi|^2-m^2|\varphi|^2-\lambda|\varphi|^4 \, .
\ee
We introduce real fields via
\be
\varphi = \frac{1}{\sqrt{2}}(\varphi_1 + i\varphi_2) \, , 
\ee
which leads to the Lagrangian
\be
{\cal L} = \frac{1}{2}\left[(\partial_0\varphi_1)^2+(\partial_0\varphi_2)^2 -(\nabla\varphi_1)^2
-(\nabla\varphi_2)^2 +2\mu(\varphi_2\partial_0\varphi_1-\varphi_1\partial_0\varphi_2) +
(\mu^2-m^2)(\varphi_1^2+\varphi_2^2)-\frac{\lambda}{2}(\varphi_1^2+\varphi_2^2)^2\right] \, .
\ee
As discussed above for the meson case, we separate the zero-mode $\phi_i$, allowing for Bose condensation,
$\varphi_i\to\varphi_i+\phi_i$. Then the Lagrangian becomes
\be
{\cal L} = -U(\phi^2) + {\cal L}^{(2)} + {\cal L}^{(3)} + {\cal L}^{(4)} , 
\ee
with 
\begin{subequations}
\bea
U(\phi^2) &=& \frac{m^2-\mu^2}{2}(\phi_1^2+\phi_2^2)  + \frac{\lambda}{4}(\phi_1^2+\phi_2^2)^2 \, , \\
{\cal L}^{(2)} &=& -\frac{1}{2}\left[-(\partial_0\varphi_1)^2-(\partial_0\varphi_2)^2 +(\nabla\varphi_1)^2 
+(\nabla\varphi_2)^2 -2\mu(\varphi_2\partial_0\varphi_1-\varphi_1\partial_0\varphi_2) \right.\non
&& \left. +\,(m^2-\mu^2)(\varphi_1^2+\varphi_2^2)+\lambda(3\phi_1^2+\phi_2^2)\varphi_1^2+
\lambda(\phi_1^2+3\phi_2^2)\varphi_2^2 + 4\lambda\phi_1\phi_2\varphi_1\varphi_2\right] \, , \\
{\cal L}^{(3)} &=& -\lambda(\phi_1\varphi_1+\phi_2\varphi_2)(\varphi_1^2+\varphi_2^2) \, , \label{Lcubic}\\
{\cal L}^{(4)} &=& -\frac{\lambda}{4}(\varphi_1^2+\varphi_2^2)^2 \, ,\label{quarticphi4}
\eea
\end{subequations}
where we have omitted terms linear in $\varphi_i$. These terms vanish upon space-time integration in the 
action:
the field $\varphi_i(X)$ does not contain a Fourier component $\varphi_i(K=0)$ because this component 
is separated and given by $\phi_i$. Therefore, the $dX$-integral over a single field 
$\varphi_i(X)\propto\sum_{K\neq 0}\exp(iK\cdot X)\varphi_i(K)$ (and
an arbitrary number of constant factors $\phi_i$) vanishes. 

\begin{figure} [ht]
\begin{center}
\includegraphics[width=0.5\textwidth]{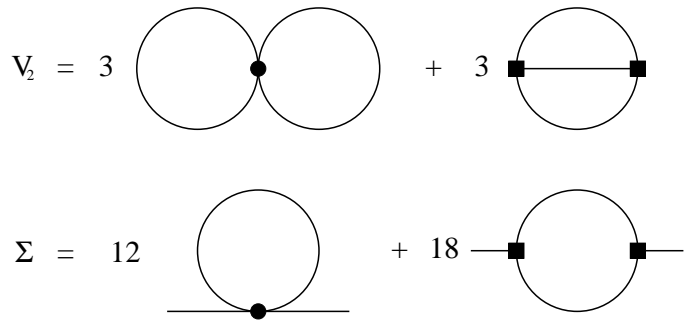}
\caption{Sum of two-particle irreducible diagrams $V_2[\phi,S]$ in the two-loop approximation
and resulting self-energy $\Sigma$. 
The black circle denotes a vertex $\lambda$ while the black squares contain 
a condensate, $\lambda\phi$. We shall neglect the second diagram in both expressions which 
originates from the induced cubic interactions.
} 
\label{figv2}
\end{center}
\end{figure}

We shall set $\phi_2$ to zero which corresponds to choosing a certain direction in the degeneracy space 
of the ground state and thus does not alter the physics. We denote $\phi\equiv\phi_1$.
The two-particle-irreducible (2PI) effective potential $V_{\rm eff}$ is a functional 
of the condensate $\phi$ and the full propagator $S$,
\be \label{Veff}
V_{\rm eff}[\phi,S] = U(\phi) + \frac{1}{2}{\rm Tr}\ln S^{-1} + 
\frac{1}{2}{\rm Tr}[S_0^{-1}(\phi)S-1] + V_2[\phi,S] \, .
\ee
Here, the trace is taken in the 2-component space of the field $(\varphi_1,\varphi_2)$ and in 
momentum space, and $S_0^{-1}(\phi)$ is the tree-level inverse propagator, in this case given 
in momentum space by the $2\times 2$ matrix 
\be \label{nondiag}
S_0^{-1} = \left(\begin{array}{cc}-K^2+m^2+3\lambda\phi^2-\mu^2 & 
-2ik_0\mu \\[1ex] 2ik_0\mu & 
-K^2+m^2+\lambda\phi^2-\mu^2 \end{array}\right) \, .
\ee
The functional $V_2[\phi,S]$ is the sum of all 2PI diagrams. 
We shall use the two-loop approximation for this
quantity. In this case, $V_2[\phi,S]$ is given by the diagrams shown in Fig.\ \ref{figv2}. Every line
in these diagrams corresponds to a full propagator $S$. The 
``double bubble diagram'' is generated by the quartic term ${\cal L}^{(4)}$ with vertex $\lambda$, represented
by a black circle. The 
diagram on the right, with two black squares,
is generated by the terms cubic in the fields 
$\varphi_1$, $\varphi_2$. In this case, every vertex contains a condensate $\phi$ 
as can be seen from ${\cal L}^{(3)}$ in Eq.\ (\ref{Lcubic}).

The ground state is given by finding the stationary point of $V_{\rm eff}$ with respect to the 
variables $\phi$ and $S$. To this end, we define the self-energy 
\be \label{defSigma}
\Sigma \equiv 2\frac{\delta V_2}{\delta S} \, .
\ee
In Fig.\ \ref{figv2}, $\Sigma$ is shown diagrammatically. It is obtained by cutting a propagator line
in each of the diagrams of $V_2[\phi,S]$. We shall neglect the right-hand
diagram in Fig.\ \ref{figv2},
which contains two vertices of the order $\lambda\phi$ (black squares). 
Then $V_2$ is a functional only of 
$S$, and not of $\phi$. This approximation is particularly simple since the self-energy becomes 
momentum-independent.

\subsection{Stationarity equations}

The self-consistent stationarity equations become 
\begin{subequations}
\bea
0 &=& \frac{\partial U}{\partial \phi} + \frac{1}{2}{\rm Tr}\left[\frac{\partial S_0^{-1}}{\partial\phi} S
\right] \, , \label{condmini}\\   
S^{-1} &=& S_0^{-1} + \Sigma \, .\label{schwinger}
\eea
\end{subequations}
The first equation arises from minimization with respect to the condensate, while the second one is the 
Dyson-Schwinger equation, obtained from minimization with respect to the propagator.
In order to evaluate these equations, we make the following ansatz for the full inverse propagator,
\be \label{ansatzS}
S^{-1}(K) = \left(\begin{array}{cc}-K^2+M_+^2-\mu^2 & 
-2ik_0\mu \\[1ex] 2ik_0\mu & 
-K^2+M^2_--\mu^2 \end{array}\right) \, ,
\ee
where $M_+$ and $M_-$ are masses that have to be determined self-consistently. 
This is the most general ansatz we can make, as we prove in Appendix \ref{App1}. 
Inversion of $S^{-1}$ yields the propagator
\be
S(K) = \frac{1}{[k_0^2-(\e_k^+)^2][k_0^2-(\e_k^-)^2]}\left(\begin{array}{cc}-K^2+M_-^2-\mu^2 & 
2ik_0\mu \\[1ex] -2ik_0\mu & 
-K^2+M_+^2-\mu^2 \end{array}\right) \, ,
\ee
where the excitation energies are given by 
\be \label{excite}
\e_k^\pm\equiv \sqrt{E_k^2+\mu^2\mp\sqrt{4\mu^2E_k^2+\delta M^4}} \, . 
\ee
We see that the sum and the difference of the two masses (squared) appear in the dispersions,
\be \label{defMdM}
E_k\equiv \sqrt{k^2+\overline{M}^2} \, , \qquad \overline{M}^2\equiv\frac{M_+^2+M_-^2}{2} \, , \qquad
\delta M^2\equiv\frac{M_+^2-M_-^2}{2} \, .
\ee
Let us abbreviate the diagonal elements of the propagator by 
\be \label{diagonalS}
S^\pm(K)\equiv \frac{-K^2+M_\mp^2-\mu^2}{[k_0^2-(\e_k^+)^2][k_0^2-(\e_k^-)^2]} \, .
\ee
Then, after dividing by $\phi$, 
the first stationarity equation becomes
\be \label{condmini1}
\mu^2=m^2+\lambda\phi^2+\lambda\frac{T}{V}\sum_Q\left[3S^+(Q)+S^-(Q)\right] \, .
\ee
For the Dyson-Schwinger equation, we need the form of the 
double-bubble diagram 
\be \label{bubble}
V_2 = 3\Lambda_{abcd}\sum_KS_{ab}(K)\sum_QS_{cd}(Q) \, , 
\ee
where the symmetrized vertex tensor follows from the form of the quartic term in the Lagrangian
${\cal L}^{(4)}$ in Eq.\ (\ref{quarticphi4}),
\be
\Lambda_{abcd}\equiv \frac{\lambda}{12}(\delta_{ab}\delta_{cd}+\delta_{ac}\delta_{bd}+\delta_{ad}\delta_{bc})
\, .
\ee
Thus, from its definition (\ref{defSigma}) we deduce the self-energy 
\be \label{self}
\Sigma = \lambda \frac{T}{V}\sum_Q \left(\begin{array}{cc} 3S^+(Q)+S^-(Q) &0 
\\ 0 & S^+(Q)+3S^-(Q) \end{array}\right) \, .
\ee
Here, the off-diagonal elements vanish due to the antisymmetry of the matrix $S$.
Consequently, with Eqs.\ (\ref{nondiag}) and (\ref{ansatzS}) 
the matrix equation (\ref{schwinger}) reduces to the two scalar equations
\begin{subequations}
\bea
M_+^2 &=& m^2+3\lambda\phi^2 + \lambda\frac{T}{V}\sum_Q [3S^+(Q)+S^-(Q)] \, , \label{schwinger1}\\
M_-^2 &=& m^2+\lambda\phi^2 + \lambda\frac{T}{V}\sum_Q [S^+(Q)+3S^-(Q)] \, . \label{schwinger2} 
\eea
\end{subequations}
The momentum sum from Eq.\ (\ref{schwinger1}) appears also in the first stationarity equation
(\ref{condmini1}). We can use this fact to simplify the latter. Moreover, we may write the equations 
in terms of the mass sum and difference appearing in the excitation energies (\ref{excite}). To this
end, we add/subtract Eqs.\ (\ref{schwinger1}) and (\ref{schwinger2}) to/from each other. 
Finally, we perform the Matsubara sum and take the thermodynamic limit 
to arrive at the set of equations (for details see Appendix \ref{appB})
\begin{subequations}\label{exact}
\bea
\overline{M}^2+\delta M^2-\mu^2&=& 2\lambda\phi^2 \, , \label{M}\\
\overline{M}^2&=&m^2+2\lambda\phi^2+2\lambda\,I \, , \label{MM}\\
\d M^2 &=& \frac{\lambda\phi^2}{1+\lambda\, J} \, ,\label{dM}
\eea
\end{subequations} 
with the momentum integrals
\bea \label{momintegrals}
I&\equiv & \sum_{e=\pm}\int\frac{d^3{\bf q}}{(2\pi)^3}
\frac{f(\e_q^e)}{\e_q^e}\left(1-e\frac{2\mu^2}{\sqrt{4\mu^2E_q^2+\delta M^4}}\right)  \, , \qquad 
J\equiv  \sum_{e=\pm} e \int\frac{d^3{\bf q}}{(2\pi)^3}
\frac{f(\e_q^e)}{\e_q^e}\frac{1}{\sqrt{4\mu^2E_q^2+\delta M^4}} \, , 
\eea
where $f(x)\equiv 1/[\exp(x/T)-1]$ is the Bose distribution function.
Expanding in the coupling $\lambda$, we may neglect the term $\lambda J$
in the denominator on the right-hand side of Eq.\ (\ref{dM}). We also approximate
\be
I\simeq \sum_{e=\pm}\int\frac{d^3{\bf q}}{(2\pi)^3}
\frac{f(\e_q^e)}{E_q} \, .
\ee
We have confirmed numerically that, for the results we present below,  
the exact result is practically indistinguishable from the approximate one. We shall see
that in the case of kaons in CFL, the quantity that plays the role of the coupling $\lambda$ 
is sufficiently small to use this approximation too.
This approximation also has the advantage that the Goldstone mode is explicitly gapless. This can be
seen by inserting Eq.\ (\ref{M}) and $\delta M^2\simeq\lambda\phi^2$ from Eq.\ (\ref{dM})  into the
excitation energy (\ref{excite}). One obtains $\e_{k=0}^+ = 0$ for all temperatures below $T_c$. 
Strictly speaking, Eqs.\ (\ref{exact}) violate the Goldstone theorem by producing a small 
energy gap for the Goldstone mode. This shortcoming of the present formalism  
has been observed in previous works in the context of linear sigma models 
\cite{Roder:2003uz,Lenaghan:1999si}. (For a self-consistent, gapless solution at nonzero temperatures
see Ref.\ \cite{Yukalov:2006wk}.)  

\subsection{Evaluation of stationarity equations}

\begin{figure} [ht]
\begin{center}
\hbox{\includegraphics[width=0.5\textwidth]{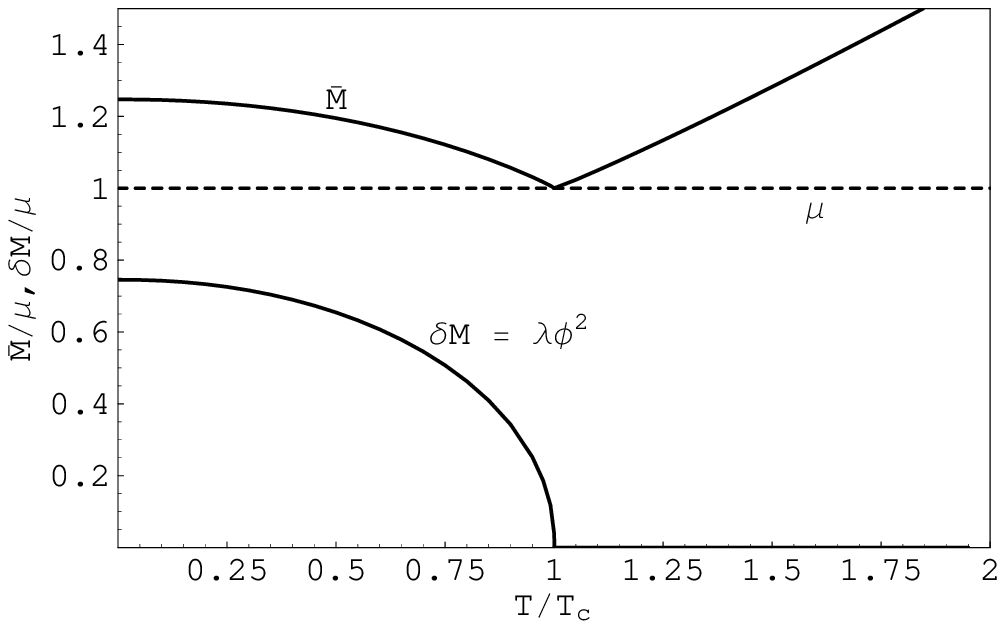}
\includegraphics[width=0.5\textwidth]{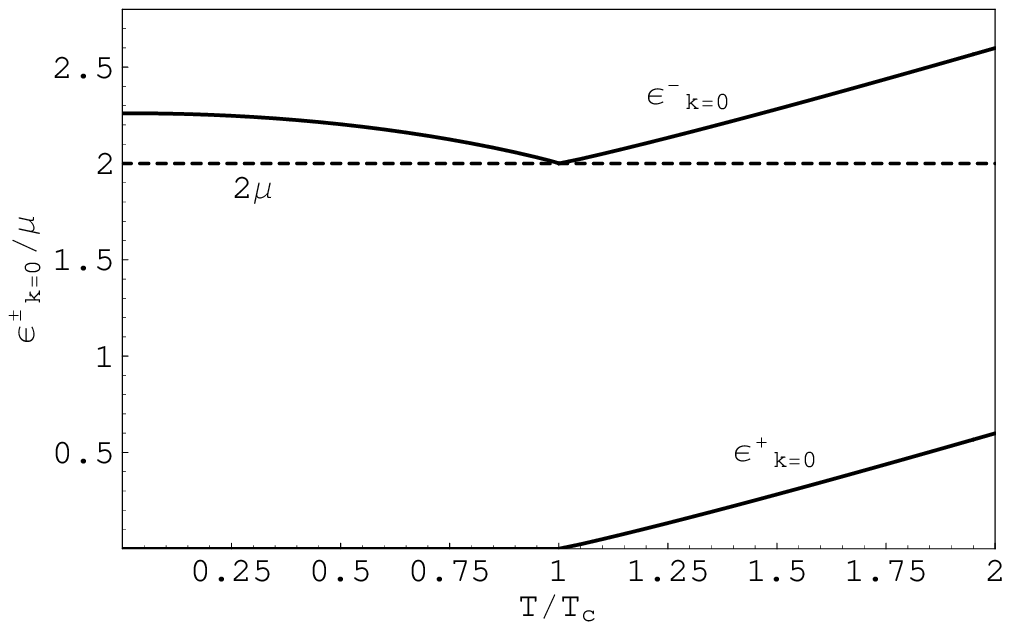}
}
\caption{
Results of the self-consistency equations (\ref{exact}) for $\varphi^4$ theory for a 
coupling constant
$\lambda=0.01$ and a chemical potential $\mu=1.5m$. Left panel: self-consistent mass 
$\overline{M}$ and $\delta M\simeq\lambda\phi^2$ (solid lines) in units of the chemical potential 
as functions of temperature. The dashed line marks, for comparison, the chemical potential. 
Right panel: energy gaps $\e_{k=0}^\pm$ in units of the chemical potential. The Goldstone mode
is gapless in the symmetry-broken phase, while the other mode assumes its minimum $2\mu$ at the critical 
point.
} 
\label{figMpMmphi4}
\end{center}
\end{figure}

At zero temperature, $I=J=0$, and Eqs.\ (\ref{exact}) yield 
\be\label{T0}
\phi^2= \frac{\mu^2-m^2}{\lambda} \, , \qquad
\overline{M}^2=2\mu^2-m^2 \, , \qquad
\d M^2=\mu^2-m^2 \, .
\ee
Nonzero temperatures have to be studied numerically. In
Fig.\ \ref{figMpMmphi4} we present the results of a numerical
evaluation for the parameters $\lambda=0.01$ and $\mu=1.5 m$. The mass
$\overline{M}$ decreases with increasing temperature until
it equals
the chemical potential at the critical point. This is expected from
Eqs.\ (\ref{exact}), which yield $\delta M=0$ and
$\overline{M}^2=\mu^2$ at $T=T_c$. This is also expected by comparison
with the theory without chemical potential. In that case we know that
the field is massless at the critical point. Here, the zero point is
replaced by the chemical potential.  At the critical point,
$\overline{M}$ is non-analytic, indicating the second order phase
transition. Above $T_c$ it increases and becomes linear in $T$ for
sufficiently large temperatures. For $T>T_c$, Eq.\ (\ref{M}) is
automatically fulfilled (we have divided by $\phi$ to obtain this
equation). Hence, above $T_c$ we simply have to solve Eq.\ (\ref{MM})
for the single variable $\overline{M}$. It is worth pointing out that
$\overline{M}>\mu$ for all temperatures. Thus, the dispersions are
positive for all momenta.  Remember that the simplest approximations
fail to fulfill this elementary requirement, see discussion below
Eq.\ (\ref{Eie}).  The mass $\delta M$ vanishes above $T_c$ since it is
proportional to the condensate $\phi$.

In the high-temperature limit we may obtain an analytic expression for $T_c$. To this end, we approximate 
the integral $I$ in Eq.\ (\ref{MM}) by $I\simeq T^2/6$. Then, the critical temperature becomes
\be
T_c^2=3\,\frac{\mu^2-m^2}{\lambda} \, .
\ee
The minimum of the two excitation branches $\e_{k=0}^\pm$ 
is shown in the right panel of Fig.\ \ref{figMpMmphi4}.
We see that one of the branches is (approximately) massless, as expected. The other branch 
has an energy gap of at least $2\mu$. In a theory without chemical potential we would expect the 
second branch to be massless too at the critical temperature. Here, this corresponds to an energy gap of
$2\mu$ at $T=T_c$.

\section{Self-consistent kaon masses in CFL matter and critical temperature}
\label{seckaons}

We are now prepared to turn to the nonzero-temperature analysis of a system of neutral and 
charged kaons in CFL quark matter, taking into account meson condensation. We shall restrict ourselves
to the case of only one condensed component. This is the most likely scenario in the case of fixed
meson chemical potentials. Only after including neutrality conditions and neutrino trapping might we 
expect a coexistence of two mesonic condensates \cite{Kaplan:2001qk}. We shall not consider this 
possibility in this paper. 

We shall apply the self-consistent method from Sec.\ \ref{secphi4}. From Sec.\ \ref{seczeroT} we 
extract the relevant terms from the effective chiral Lagrangian that reduce this theory
essentially to a $\varphi^4$ theory. We shall see that basic parameters of the chiral effective theory
such as the meson masses and chemical potentials play, in certain combinations, the role of the 
coupling constant $\lambda$ of $\varphi^4$ theory. The most important difference to $\varphi^4$ theory
is the presence of a second, uncondensed boson. Technically, this does not create any additional
difficulties. Therefore, we can directly set up the self-consistent set of equations upon referring 
to the previous subsection. Then, we shall  solve them and discuss the physical results, in particular the 
value of the critical temperature.

\subsection{Derivation of self-consistency equations}
\label{deriveq}

We expand the potential $U$ from Eq.\ (\ref{Uallorder}) up to fourth order in the condensates,
\bea \label{Ufourthorder}
U(\phi_1,\phi_2)  &\simeq& \frac{m_1^2-\mu_1^2}{2}\phi_1^2+\frac{m_2^2-\mu_2^2}{2}\phi_2^2
+\frac{\beta_1}{4}\phi_1^4+\frac{\beta_2}{4}\phi_2^4+\frac{\alpha}{2}\phi_1^2\phi_2^2 \, .
\eea
For notational convenience, here and in the following we use the subscript 1 for $K^+$ and 2 for $K^0$.
We have abbreviated  
\be \label{alphabeta}
\beta_{i}\equiv\frac{1}{6f_\pi^2}(4\mu^2_{i}-m^2_{i}) \, , \qquad \alpha\equiv \frac{1}{12f_\pi^2}
(\mu_1^2+\mu_2^2-m_1^2-m_2^2+6\mu_1\mu_2) 
=\frac{\beta_1+\beta_2}{2}-\left(\frac{\mu_1-\mu_2}{2f_\pi}\right)^2 \, .
\ee
These dimensionless quantities play the role of coupling constants. They can be positive or negative,
depending on the bare meson masses and chemical potentials . If the 
condensates were allowed to become arbitrarily large, boundedness of the potential (\ref{Ufourthorder})
from below would require $\beta_1,\beta_2>0$ and $\alpha>-(\beta_1\beta_2)^{1/2}$. However, 
here we consider only small condensates, since we have employed an expansion of   
the full potential (\ref{Uallorder}) (the full potential is bounded for all possible parameters).
Therefore, we do not have any restrictions for the coupling constants.

In order to determine the $4\times 4$ tree-level propagator for charged and neutral kaons, we consider 
terms quadratic in the field $\theta$, Eqs.\ (\ref{quad}). We perform the trace in these terms, 
set the $K^+$ condensate to zero, and denote the $K^0$ condensate by $\phi$. Then, the 
inverse tree-level propagator is block diagonal,
\be \label{treeprop}
S_0^{-1} = \left(\begin{array}{cc}S_{01}^{-1} & 0 \\ 0 & S_{02}^{-1} \end{array}\right) \,  , 
\ee
where
\begin{subequations}
\bea
S_{01}^{-1} &=& \left(\begin{array}{cc} -k_0^2+v_\pi^2k^2 + m_1^2-\mu_1^2+\alpha\phi^2  & 
-2i\mu_1k_0 \\[1ex] 2i\mu_1k_0 &
-k_0^2+v_\pi^2k^2 + m_1^2-\mu_1^2+\alpha\phi^2  \end{array}\right) \, , \\ \non
S_{02}^{-1} &=& \left(\begin{array}{cc} -k_0^2+v_\pi^2k^2 + m_2^2-\mu_2^2 +3\beta_2\phi^2  & 
-2i\mu_2k_0 \\[1ex] 
2i\mu_2k_0 &
-k_0^2+v_\pi^2k^2 + m_2^2-\mu_2^2 +\beta_2\phi^2 \end{array}\right) \, .
\eea
\end{subequations}
For $\phi=0$ we recover the ideal gas approximation of the propagator, Eq.\ (\ref{Sideal}).
Upon comparing the inverse propagator of the condensed branch, $S_{02}^{-1}$, with the one 
from $\varphi^4$ theory, Eq.\ (\ref{nondiag}), we see that $\beta_2$ plays the role
of the coupling constant $\lambda$. Note that the condensate also gives a correction to the bare mass
of the uncondensed branch. In this branch, $\alpha$ assumes the role of $\lambda$ 
(with the important difference, however, that the correction is the same for both degrees of freedom).
 
Analogous to above, we use the following ansatz for the full inverse propagator,
\be
S^{-1} = \left(\begin{array}{cc}S_{1}^{-1} & 0 \\ 0 & S_{2}^{-1} \end{array}\right) \,  , 
\ee
where 
\bea
S_{i}^{-1} &=& \left(\begin{array}{cc} -k_0^2+v_\pi^2k^2 + M_{i,+}^2-\mu_{i}^2  & 
-2i\mu_{i}k_0 \\[1ex] 2i\mu_{i}k_0 &
-k_0^2+v_\pi^2k^2 + M_{i,-}^2-\mu_{i}^2  \end{array}\right) \, . 
\eea
We have introduced the four self-consistent masses $M_{1,\pm}$, $M_{2,\pm}$. 
Not surprisingly, we shall see below that
for the uncondensed branch both masses in fact coincide, $M_{1,+}=M_{1,-}$. Inversion yields the 
propagator
\be \label{fullprop}
S = \left(\begin{array}{cc}S_{1}& 0 \\ 0 & S_{2} 
\end{array}\right) \,  , 
\ee
where
\be
S_{i} = \frac{1}{[k_0^2-(\e_{i}^+)^2][k_0^2-(\e_{i}^-)^2]}\left(\begin{array}{cc}-k_0^2+v_\pi^2k^2+M_{i,-}^2
-\mu_{i}^2 & 
2ik_0\mu_{i} \\[1ex] -2ik_0\mu_{i} & 
-k_0^2+v_\pi^2k^2+M_{i,+}^2-\mu_{i}^2 \end{array}\right) \, .
\ee
We have defined the excitation energies 
\be \label{excitekaon}
\e_{i}^\pm\equiv \sqrt{E_{i}^2+\mu_{i}^2\mp\sqrt{4\mu_{i}^2E_{i}^2+\delta M_{i}^4}}
 \, , 
\ee
with 
\be \label{defMdMkaon}
E_{i}\equiv \sqrt{v_\pi^2k^2+\overline{M}_{i}^2} \, , \qquad \overline{M}_{i}^2
\equiv\frac{M_{i,+}^2+M_{i,-}^2}{2} \, , \qquad
\delta M_{i}^2\equiv\frac{M_{i,+}^2-M_{i,-}^2}{2} \, .
\ee
For $\delta M=0$, one recovers the form of the excitation energy $\e_{i}^\pm=E_{i}\mp \mu_{i}$. 
In the following, we shall abbreviate the diagonal elements of the kaon propagator by
\be
S_{i}^\pm\equiv \frac{-k_0^2+v_\pi^2k^2 + M_{i,\mp}^2-\mu_{i}^2}
{[k_0^2-(\e_{i}^+)^2][k_0^2-(\e_{i}^-)^2]} \, .
\ee   
For the stationarity equation (\ref{condmini}) we set $\phi_1=0$ and $\phi_2=\phi$ in the 
potential $U$ from Eq.\ (\ref{Ufourthorder}) to obtain for nonzero $\phi$
\bea  \label{stat1kaon}
\mu_2^2&=&m_2^2+\beta_2\phi^2+\frac{T}{V}\sum_Q\left[\alpha(S_{1}^++S_{1}^-)
+\beta_2(3S_{2}^++S_{2}^-)\right] \, .
\eea

\begin{figure} [ht]
\begin{center}
\includegraphics[width=0.9\textwidth]{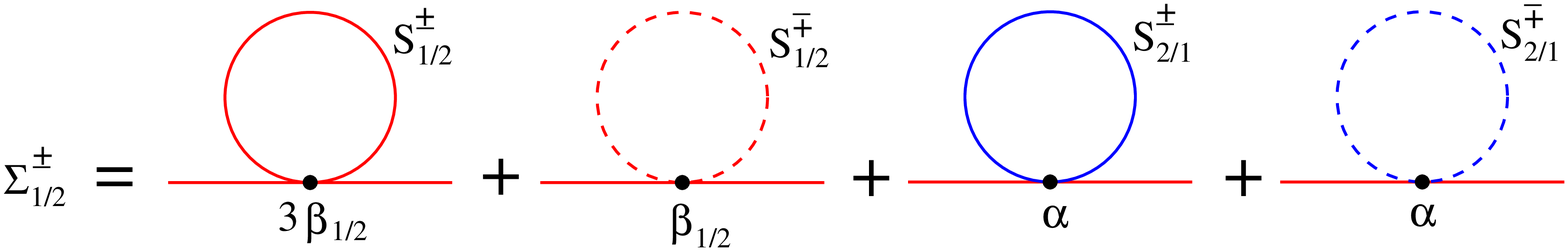}
\caption{
(Color online) Kaon self-energy $\Sigma_{1/2}^\pm$ (meaning
$\Sigma_1^\pm$ {\em or} $\Sigma_2^\pm$) according to
Eq.\ (\ref{lambda1p}). The different contributions originate from
interactions with the same kaon species (red propagators, same
subscript) and with the opposite species (blue propagators, opposite
subscript). The effective couplings for these interactions are
$\beta_1$ (interaction between charged kaons),
$\beta_2$ (interaction between neutral kaons) 
and $\alpha$ (interaction between charged and neutral kaon), see definition
(\ref{alphabeta}). Moreover, in the case of self-coupling, there is
a difference between coupling to the same degree of freedom (solid
propagators) and the orthogonal degree of freedom (dashed
propagators), the former being larger by a factor 3 than the latter.
}
\label{figsigma}
\end{center}
\end{figure}

Next, we need the self-energy $\Sigma$. The double-bubble diagram has the same form as in 
Eq.\ (\ref{bubble}) (now of course with $a,b,c,d$ running from 4 through 7), where the symmetrized
vertex tensor is extracted from the quartic term in the Lagrangian, Eq.\ (\ref{L21}),
\be 
\Lambda_{abcd} = -\frac{1}{4!}\frac{1}{2f_\pi^2}\left\{
{\rm Tr}\left[\frac{1}{3} AT_a AT_bT_cT_d-\frac{1}{12}XT_aT_bT_cT_d-
\frac{1}{4}AT_aT_bAT_cT_d\right] + \mbox{all permutations of $a,b,c,d$}\right\} \, .
\ee
The self-energy is
\be
\Sigma_{ab} = 12\Lambda_{abcd}\sum_Q S_{cd}(Q) \, .
\ee
Since the propagator $S$ is antisymmetric, the off-diagonal elements do not contribute
in the contraction with the totally symmetric tensor $\Lambda_{abcd}$. 
Consequently, $\Sigma$ only contains the diagonal elements $S_{i}^\pm$. 
Performing the traces and contractions yields
\be \label{lambda1p}
\Sigma \equiv {\rm diag}(\Sigma_1^+,\Sigma_1^-,\Sigma_2^+,\Sigma_2^-) = 
\frac{T}{V}\sum_Q (\lambda_1^+ S_1^++\lambda_1^- S_1^-+\lambda_2^+ S_2^++\lambda_2^- S_2^-)
\, ,
\ee
with
\begin{subequations}
\bea
\lambda_1^+ &=& {\rm diag}(3\beta_1,\beta_1,\alpha,\alpha) \, , \qquad 
\lambda_1^- = {\rm diag}(\beta_1,3\beta_1,\alpha,\alpha)\, , \\ 
\lambda_2^+ &=& {\rm diag}(\alpha,\alpha,3\beta_2,\beta_2) \, , \qquad 
\lambda_2^-= {\rm diag}(\alpha,\alpha,\beta_2,3\beta_2) \, .
\eea
\end{subequations}
These diagonal matrices yield the effective coupling constants for the respective contributions
to the self-energy. We illustrate these contributions in Fig.\ \ref{figsigma}. An interaction between kaons
of the same species is given by the effective coupling $\beta_i$, while an interaction
between kaons of different species is given by $\alpha$. Condensation of species
$i$ implies $\mu_i>m_i$ and thus a positive ``self-coupling'' $\beta_i$. The self-coupling of 
the uncondensed species, however, as well as the coupling $\alpha$ can become negative 
for certain values of the bare masses and chemical potentials. 

With the self-energy (\ref{lambda1p}), the Dyson-Schwinger equation (\ref{schwinger}) 
becomes the set of four equations
\begin{subequations}
\bea
M_{1,+}^2&=&m_1^2+\alpha\phi^2+\frac{T}{V}\sum_Q \left[\alpha(S_{2}^++S_{2}^-)+\beta_{1}(3S_{1}^++S_{1}^-)
\right] \, , \label{one}\\
M_{1,-}^2&=&m_1^2+\alpha\phi^2+\frac{T}{V}\sum_Q \left[\alpha(S_{2}^++S_{2}^-)+\beta_{1}(S_{1}^++3S_{1}^-)
\right] \, , \label{two}\\
M_{2,+}^2&=&m_2^2+3\beta_2\phi^2+\frac{T}{V}\sum_Q \left[\alpha(S_{1}^++S_{1}^-)+\beta_{2}(3S_{2}^++S_{2}^-)
\right] \, , \label{three}\\
M_{2,-}^2&=&m_2^2+\beta_2\phi^2+\frac{T}{V}\sum_Q \left[\alpha(S_{1}^++S_{1}^-)+\beta_{2}(S_{2}^++3S_{2}^-)
\right] \, .\label{four}
\eea
\end{subequations}
We now insert Eq.\ (\ref{three}) into the first stationarity condition (\ref{stat1kaon}). Furthermore, 
we add/subtract Eqs.\ (\ref{three}) and (\ref{four}) to/from each other. Finally, from subtracting
Eq.\ (\ref{two}) from Eq.\ (\ref{one}) we conclude $M_{1,+}=M_{1,-}=\overline{M}_1$ such that we can drop
Eq.\ (\ref{two}). 
Then, after performing the Matsubara sum, we arrive at the following set of four equations
for the four variables $\phi$, $\overline{M}_1$, $\overline{M}_2$, $\delta M_2$,
\begin{subequations} \label{foureqs}
\bea
\overline{M}^2_2-\mu_2^2 &=& \beta_2\phi^2 \, , \label{eq1} \\
\overline{M}_1^2&=&m_1^2+\alpha\phi^2+\alpha I_2+2\beta_1 I_1 \, , \label{eq2} \\
\overline{M}^2_2&=&m_2^2+2\beta_2\phi^2+\alpha I_1+2\beta_2 I_2 \, , \label{eq3} \\
\delta M_2^2 &=& \beta_2\phi^2 \, , \label{eq4}
\eea
\end{subequations}
with the momentum integrals
\be \label{defIi}
I_{i}\equiv\frac{T}{V}\sum_Q[S_{i}^+(Q)+S_{i}^-(Q)]\simeq\sum_{e=\pm}\int\frac{d^3{\bf q}}{(2\pi)^3}
\frac{f(\e_{i}^e)}{E_{i}} \, .
\ee
We have made use of the same approximations as explained below Eq.\ (\ref{momintegrals}) in the context of 
$\varphi^4$ theory. In this approximation, one of the branches of the condensed species is 
explicitly gapless, $\e_2^+(k=0)=0$ for all temperatures smaller than the critical temperature.

\subsection{Solutions for zero and nonzero temperature: melting the kaon condensate}
\label{solveeq}

We may now solve the system of equations (\ref{foureqs}). In these equations, the kaon masses and 
chemical potentials are kept as parameters. A numerical evaluation should be done for values 
of these parameters that are realistic for the interior of a compact star. However, both masses and 
chemical potentials are known only with significant uncertainties. These uncertainties
originate partly from the matching calculations that determine the decay constant $f_\pi$ and the 
constant $a$, see Eq.\ (\ref{matching}). These calculations have been performed using
perturbative methods within QCD. Hence they are reliable only for quark chemical potentials 
far beyond realistic values. Extrapolations from these
high-density values yield $f_\pi\simeq 100$ MeV and $a\simeq 0.01$ for $\mu\simeq 500$ MeV and 
$\Delta\simeq 30$ MeV. The next source of uncertainties are the quark masses that enter the 
expressions of the meson masses and chemical potentials, see Eq.\ (\ref{masschem}). Using a strange mass of 
$m_s\simeq 150$ MeV, light quark masses $m_d\simeq 7$ MeV, $m_u\simeq 4$ MeV, and  
$\mu_Q=0$ we find that the kaon chemical potentials are of the order of tens of MeV, 
$\mu_1\simeq\mu_2 \simeq 20$ MeV, while their zero-temperature 
masses are of the order of a few MeV, $m_1\simeq 5$ MeV, $m_2\simeq 4$ MeV. Due to the slightly larger 
charged kaon mass, it is reasonable to assume condensation of the neutral kaon, see zero-temperature 
phase diagram in Fig.\ \ref{figmu1mu20}. We have also neglected instanton effects which yield an additional 
correction to the masses. Its value, however, is poorly known, since high-density calculations yield 
results that strongly depend on the QCD scale $\Lambda_{\rm QCD}$. The value of these corrections 
has been estimated to be of the order of 10 MeV \cite{Schafer:2002ty}. 

In our numerical results, we shall set the values of the kaon masses to the above values. As for the chemical
potentials, we have to recall our expansion in the condensate. This expansion is valid for 
small condensates and thus for small values of $1-m_2^2/\mu_2^2$. In other words, the chemical 
potential has to be larger than the mass (otherwise there is no condensation) -- this is likely to 
be the case judging from the above estimates. However, it should 
be only slightly larger (otherwise
our approximation is not reliable) -- this is questionable but not excluded from the above estimates. 
Further studies which take into account all orders in the 
condensate are required to confirm our estimates for the cases where $1-m_2^2/\mu_2^2$ is not small.
Finally, we have to recall that the effective theory itself is only applicable for energies smaller
than the fermionic energy gap $\Delta$. 
Also, the critical temperature $T_c^{\rm CFL}$ for the transition from 
CFL to unpaired quark matter of course sets an upper limit for our treatment. This temperature is 
of the order of the zero-temperature gap, $T_c^{\rm CFL}=0.57\cdot 2^{1/3}\Delta \simeq 0.7\Delta$
according to BCS theory (gauge field fluctuations, however, increase the critical temperature 
\cite{Giannakis:2004xt}.) Hence, 
when we find critical temperatures $T_c$ for kaon condensation 
on the order of or larger than $\Delta$ and $T_c^{\rm CFL}$, the conclusion 
is ``the critical temperature is in a 
regime where we do not trust the effective theory'' rather than 
``the critical temperature is $T_c$''.  
These are the premises under which our results have to be understood. 

Let us first discuss the simple case of zero temperature, where $I_{i}$ in Eqs.\ (\ref{foureqs})
vanishes. In this case, 
\begin{subequations}\label{zerotemp}
\bea
\phi^2&=&\frac{\mu_2^2-m_2^2}{\beta_2} \simeq 2f_\pi^2\left(1-\frac{m_2^2}{\mu_2^2}\right)\, , 
\label{phizeroT} \\
\overline{M}_{1}^2&=&m_1^2+\frac{\alpha}{\beta_2}(\mu_2^2-m_2^2) \, , \label{uncondensed} \\
\overline{M}_{2}^2&=&2\mu_2^2-m_2^2 \, ,\label{condensed}\\
\delta M^2_2&=&\mu_2^2-m_2^2 \, .
\eea
\end{subequations}
In Eq.\ (\ref{phizeroT}) we have expanded the condensate $\phi^2$ with respect to the smallness parameter 
$1-m_2^2/\mu_2^2$ and used the definitions for $\alpha$ and $\beta_2$ from Eq.\ (\ref{alphabeta}).
The result is in agreement 
with the expansion of the exact result (\ref{costheta}). 
We may compare the zero-temperature masses with the bare masses. For the condensed 
meson the self-consistent mass is larger, $\overline{M}_{2}>m_2$. For the uncondensed meson, this is 
not necessarily true since the correction 
through the condensate in Eq.\ (\ref{uncondensed}) becomes negative if $\alpha<0$. 
We also see that the uncondensed kaon becomes gapless in the case of equal masses and chemical 
potentials, $m_1=m_2$, $\mu_1=\mu_2$. In this case, we conclude from Eq.\ (\ref{alphabeta}) that 
$\alpha=\beta_1=\beta_2$ and thus $\overline{M}_1^2=\mu_1^2$. Consequently, $\e_1^+(k=0)=0$. 
This reflects the fact that exact isospin symmetry, $m_u=m_d$ and thus $m_1=m_2$ and $\mu_1=\mu_2$, 
together with (neutral) kaon condensation leads to two exact Goldstone modes \cite{Miransky:2001tw}. 

Let us now turn to nonzero temperatures. Before solving the equations for arbitrary temperatures, we may 
first derive an approximate expression for the critical temperature. We do so for temperatures 
large compared to the kaon masses and chemical potentials. 
We set $\phi=0$ in Eqs.\ (\ref{foureqs}) and approximate 
\be \label{Iapprox}
I_1\simeq I_2\simeq \frac{T^2}{6v_\pi^3} \, .
\ee
Then, we obtain for the critical temperature 
\bea
T_c^2 &=& 6v_\pi^3\frac{\mu_2^2-m_2^2}{\alpha+2\beta_2} = \frac{72v_\pi^3 f_\pi^2(\mu_2^2-m_2^2)}
{5(\mu_2^2-m_2^2)+(\mu_1^2-m_1^2)+6\mu_2(\mu_1+2\mu_2)} \, . \label{Tcapprox}
\eea
where we used the definitions of $\alpha$ and $\beta_{2}$ in Eq.\ (\ref{alphabeta}).
The masses at $T=T_c$ are 
\bea
\overline{M}_1^2 &=& m_1^2+(\mu_2^2-m_2^2)\frac{\alpha+2\beta_1}{\alpha+2\beta_2} \, , \qquad
\overline{M}_2^2 = \mu_2^2 \, .
\eea
As it turns out, these high-temperature expressions are a very good approximation to our numerical results.
For the critical temperature results we shall present below, the approximation is 
practically indistinguishable from the full numerical evaluation. 
From Eq.\ (\ref{Tcapprox}) we see that for $\alpha+2\beta_2<0$ there is no solution for the 
critical temperature (of course assuming that $T_c$ is real). We comment on this apparently unphysical
result in Appendix \ref{AppC}.

\begin{figure} [ht]
\begin{center}
\hbox{\includegraphics[width=0.5\textwidth]{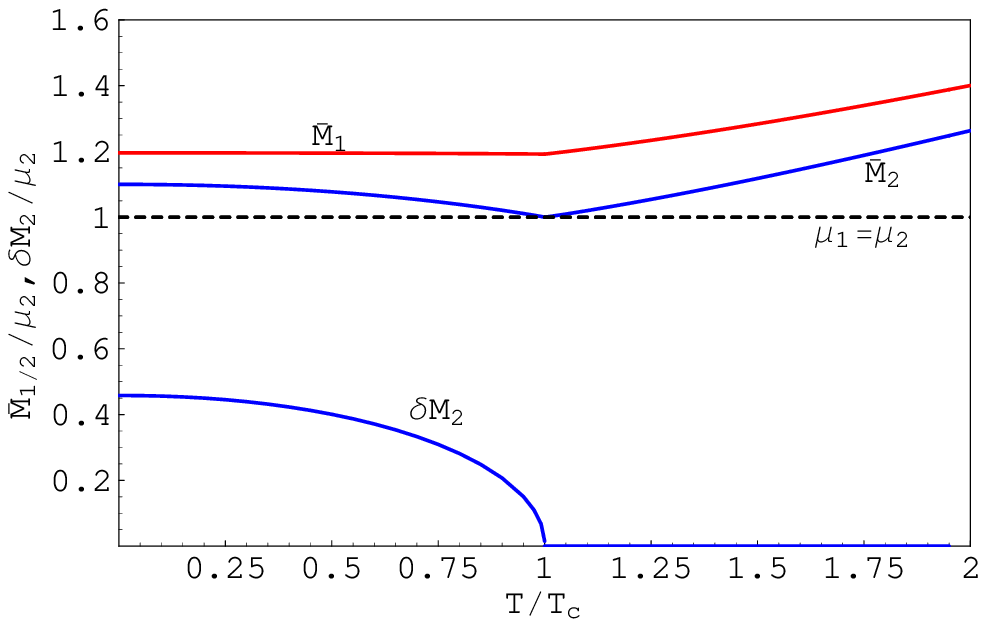}
\includegraphics[width=0.5\textwidth]{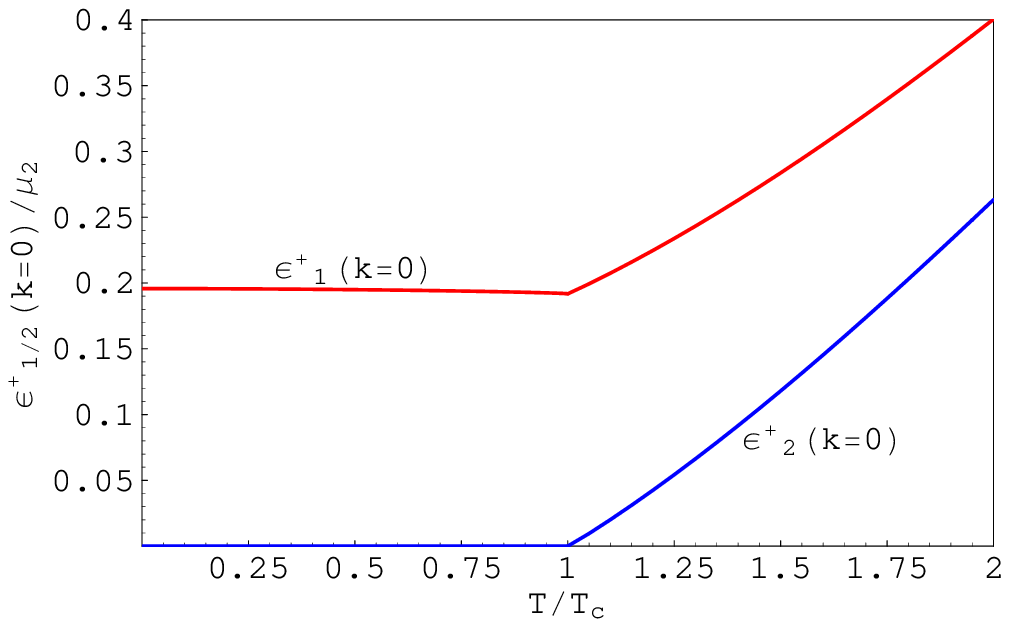}
}
\caption{
(Color online) Self-consistent kaon masses and energy gaps as a function of temperature for 
$m_1=5$ MeV, $m_2=4$ MeV,  $\mu_1=\mu_2=4.5$ MeV. 
The critical temperature in this case is $T_c\simeq 40$ MeV.
Left panel: mass of the 
condensed neutral kaon $\overline{M}_{2}$ (upper blue solid curve), and $\delta M_2\propto\phi$ 
(lower blue solid curve) as well as mass 
of the uncondensed charged kaon $\overline{M}_{1}$ (red solid curve)
in units of the chemical potential of both kaons, here chosen to be equal (and shown as the black 
dashed line). Right panel: 
energy gaps of the modes $\e_1^+$ and $\e_2^+$
(the energy gaps $\e_i^-$
are larger and not shown in the figure, cf.~similar plot in Fig.\ \ref{figMpMmphi4}). The 
condensed mode (blue curve) is gapless below the critical temperature, in accordance with the Goldstone 
theorem.   
} 
\label{figsmallmu1}
\end{center}
\end{figure}

\begin{figure} [ht]
\begin{center}
\hbox{\includegraphics[width=0.5\textwidth]{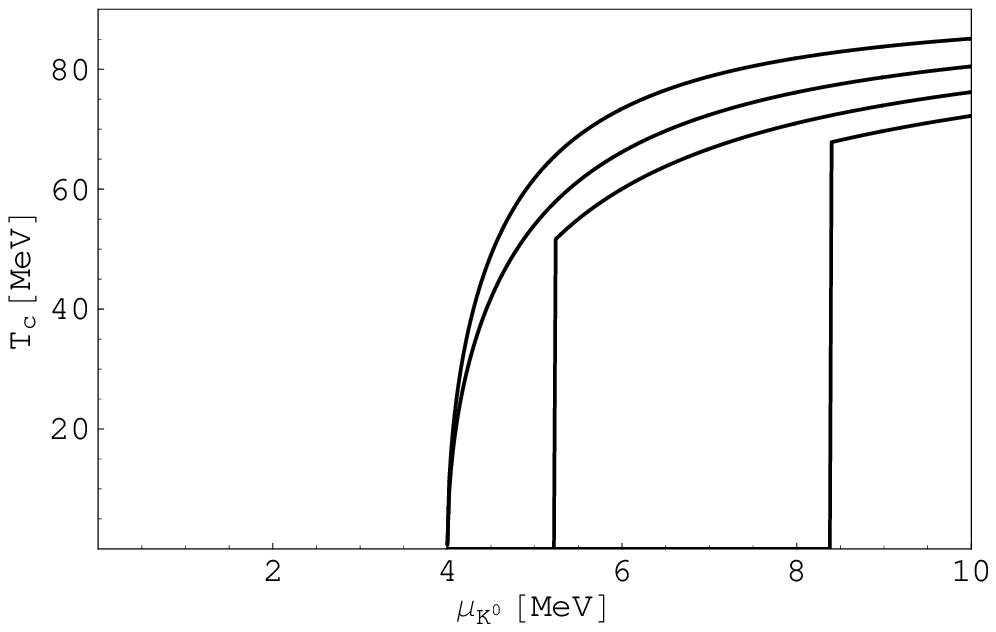}
\includegraphics[width=0.5\textwidth]{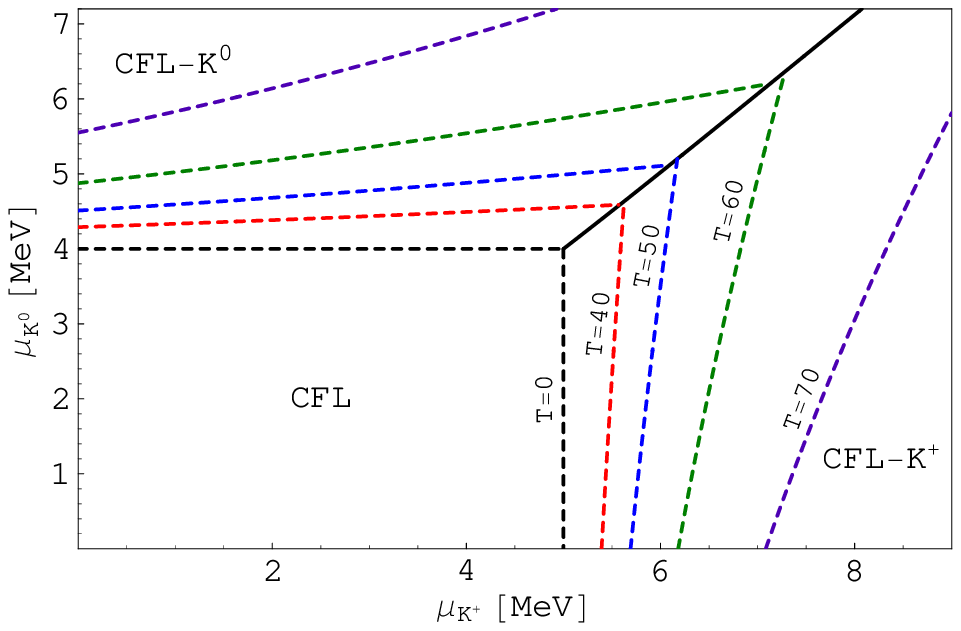}
}
\caption{(Color online) 
Critical temperature and nonzero-temperature phase diagram for bare kaon masses as 
in Fig.\ \ref{figsmallmu1},  $m_{K^+}=5$ MeV, $m_{K^0}=4$ MeV.
Left panel: $T_c$ in units of MeV as a function of $\mu_{K^0}$ for fixed charged kaon chemical 
potentials $\mu_{K^+}=0.2,3.2,6.2,9.2$ MeV from top to bottom. Right panel: 
nonzero-temperature phase diagram in the $\mu_{K^+}$-$\mu_{K^0}$-plane.
From left to right and bottom to top (in the order black, red, blue, green, violet) the phase transition 
lines correspond to $T=0,40,50,60,70$ MeV. 
} 
\label{figmu1mu2T}
\end{center}
\end{figure}

We now solve Eqs.\ (\ref{foureqs}) numerically from $T=0$ up to $T=2T_c$.
In Fig.\ \ref{figsmallmu1} we present the self-consistent kaon masses and energy gaps for 
bare masses $m_1=5$ MeV, $m_2=4$ MeV, and chemical potentials $\mu_1=\mu_2=4.5$ MeV. As can be
seen from the phase diagram in Fig.\ \ref{figmu1mu20}, this corresponds to a $K^0$ condensed
phase at $T=0$. We find that the critical temperature is $T\simeq 40$ MeV, and the condensate at 
zero temperature $\phi/f_\pi\simeq 0.6$. Since our expansion in the condensate requires
$\phi/f_\pi\ll 1$, this parameter region is close to the limit above which we do not 
trust the results quantitatively. The qualitative behavior of the quantities shown in Fig.\ \ref{figsmallmu1}
is easy to understand in view of the results of $\varphi^4$ theory in the previous section. 
The uncondensed kaon, absent in the scalar $\varphi^4$ theory, exhibits almost constant masses 
and energy gaps below $T_c$. Its energy gap is nonzero since we have chosen the charged and
neutral kaon masses to be different, $m_1\neq m_2$. As mentioned above, 
if isospin is an exact symmetry, then both energy gaps $\e_1^+(k=0)$ and 
$\e_2^+(k=0)$ vanish. At $T=0$, this is obvious from Eqs.\ (\ref{zerotemp}). At nonzero $T$, 
we have confirmed numerically that in this symmetric case 
both energy gaps vanish for all temperatures below $T_c$.  

We finally extract the critical temperature and extend the phase diagram from 
Fig.\ \ref{figmu1mu20} to nonzero temperatures. 
In Fig.\ \ref{figmu1mu2T} we show the critical temperature as a function
of the neutral charged kaon chemical potential for several fixed charged kaon chemical potentials. 
Two of the curves start at the neutral kaon mass. 
This is clear since there is
no condensation for chemical potentials below this mass. The other two curves for $\mu_{K^+}>m_{K^+}$ 
start at a larger value, given by the relation (\ref{curve}). Below this 
critical value there is a $K^+$ condensate, but no $K^0$ condensate 
(the plot only shows the $K^0$ critical temperature).
The jump of $T_c$ at this critical chemical potential indicates a first order phase transition. 
  
The right panel shows the phase transition lines for both $K^0$ and $K^+$ condensates. 
Here, the case of a charged kaon condensate is obtained from the above formulas simply by exchanging 
the indices 1 and 2. The phase transition lines are shown for temperatures between 
$T=0$ and $T=70$ MeV. Due to the melting of the kaon condensates the region of the CFL phase becomes
larger with increasing temperature. In contrast to Fig.\ \ref{figmu1mu20}, ``CFL'' does not 
mean absence of kaons because at nonzero 
temperature a thermal kaon population is present. The phase transition lines indicate transitions
from a kaon-condensed state to a state without kaon condensation or from a charged kaon 
condensate to a neutral kaon condensate. Transitions from one condensate to another are of first 
order (given by the solid line). It is interesting that the transition from a charged kaon 
condensate to the CFL phase is always of second order (second order phase transitions are given by 
dashed lines). The transition from the neutral kaon condensate to the CFL phase, however, can 
be either of first or second order. It is of first order where this transition occurs on the diagonal 
line -- more precisely, the line given by Eq.\ (\ref{curve}).
Otherwise, it is of second order. This is equivalent to the statement that for instance the two (green) 
lines corresponding to $T=60$ MeV do not meet the line $\mu_1(\mu_2^2-m_2^2)=\mu_2(\mu_1^2-m_1^2)$ at 
the same point. The do so only for equal meson masses $m_{K^+}=m_{K^0}$. Hence, the asymmetry 
in the order of phase transitions is created by the larger mass of the charged kaon.

Since we have not imposed any neutrality constraint in this section (we will do so in the next section),
most of the 
states in the phase diagram have an overall charge. Electric neutrality 
fixes the charge chemical potential $\mu_Q$. Thus, if one makes use of the kaon chemical potentials
in Eq.\ (\ref{masschem}), the neutrality condition fixes, for any $T$, a certain point in the 
phase diagram (for a fixed quark chemical potential $\mu$).

Here, ignoring the neutrality constraint, 
we obtain an estimate for the critical temperature for arbitrary values of the kaon
chemical potentials. 
The critical temperature rises rapidly with the chemical potential.
In our numerical example ($m_{K^+}=5$ MeV, $m_{K^0}=4$ MeV), when the chemical potential is only
about 10\% larger than the boson mass ($\mu_{K^0}=4.5$ MeV), 
the critical temperature has already risen to 40~MeV. If both kaon chemical
potentials are about twice the kaon mass ($\mu_{K^+}\simeq\mu_{K^0}\simeq 8$ MeV) then the critical 
temperature is about 70~MeV.
Of course, as mentioned above, the reliability of the effective theory
at these large temperatures is questionable since fermionic degrees of
freedom start to become important or the CFL phase itself has already melted.

\section{Effects of electric neutrality}
\label{secneutral}

In the previous sections we have not included any neutrality constraint. Some of the systems we have 
considered
have a nonzero overall electric charge. For instance, in the zero-temperature phase diagram in 
Fig.\ \ref{figmu1mu20}, all states with $K^+$ condensate are positively charged, 
while all others (the $K^0$-condensed states as well as the pure CFL state) are neutral. In the case of
nonzero temperatures, all states we have considered are charged due to a thermal population of charged 
kaons. In this section, we impose the constraint of overall electric neutrality and include contributions
from electrons (and positrons). Our main interest is the 
effect of the neutrality constraint on the critical temperature. 

\subsection{Kaon densities}

In order to include electrical neutrality we have to compute the kaon number densities. In fact, we only need
the density for the charged kaons, but in the following we shall provide the densities of both 
charged and neutral kaons.
To this end, we have to compute the (negative of the) 
derivative of the effective action $V_{\rm eff}$ with respect to the kaon chemical potentials $\mu_1$,
$\mu_2$ and 
evaluate the result at the stationary point. The effective potential depends on the chemical potentials 
explicitly, as well as implicitly through $\phi$ and $S$. However, at the stationary point 
the derivatives of $V_{\rm eff}$ with respect to $\phi$ and $S$ vanish by definition. Therefore, the 
densities can be computed from taking the explicit derivative of $V_{\rm eff}$ with respect to $\mu_1$,
$\mu_2$. From the effective potential given in Eq.\ (\ref{Veff}) we thus conclude 
\be \label{density1}
n_i = -\frac{\partial U}{\partial \mu_i}-\frac{1}{2}{\rm Tr}\left[\frac{\partial S_0^{-1}}{\partial \mu_i}
S\right] \, . 
\ee
Here, $U$ is obtained from Eq.\ (\ref{Ufourthorder}) by setting $\phi_1=0$ and defining $\phi\equiv \phi_2$. 
The inverse tree-level propagator $S_0^{-1}$ and the full propagator $S$ are defined in  
Eqs.\ (\ref{treeprop}) and (\ref{fullprop}), respectively. Note that the form 
of the density (\ref{density1}) does not depend on the specific approximation of the set of 2PI 
diagrams $V_2$. We have only used the fact that $V_2$ depends on the chemical potentials 
implicitly through $\phi$ and $S$, but not explicitly. 

Now we can straightforwardly compute the densities. For $n_1$ we use
\bea
-\frac{1}{2}{\rm Tr}\left[\frac{\partial S_0^{-1}}{\partial \mu_1}
S\right]  &=& \frac{T}{V}\sum_K\frac{2\mu_1(k_0^2+E_1^2-\mu_1^2)}
{[k_0^2-(\epsilon_1^+)^2][k_0^2-(\epsilon_1^-)^2]}-
\frac{1}{2}\frac{\partial \alpha}{\partial \mu_1}\phi^2\frac{T}{V}\sum_K [S_1^+(K)+S_1^-(K)]\non
&=&\frac{1}{2}\sum_e e \int\frac{d^3k}{(2\pi)^3}
\coth\frac{\epsilon_1^e}{2T} - \frac{\phi^2}{2}
\frac{\partial \alpha}{\partial \mu_1} I_1\, ,
\eea
where, in the second step, we have performed the Matsubara sum. For $n_2$ we need 
\bea \label{deriv2}
-\frac{1}{2}{\rm Tr}\left[\frac{\partial S_0^{-1}}{\partial \mu_2}
S\right] &=& \frac{T}{V}\sum_K\frac{2\mu_2(k_0^2+E_2^2-\mu_2^2)}
{[k_0^2-(\epsilon_2^+)^2][k_0^2-(\epsilon_2^-)^2]} \non
&&-\, \frac{1}{2}\frac{\partial \alpha}{\partial \mu_2}\phi^2\frac{T}{V}\sum_K [S_1^+(K)+S_1^-(K)] 
- \frac{1}{2}\frac{\partial \beta_2}{\partial \mu_2}\phi^2
\frac{T}{V}\sum_K [3S_2^+(K)+S_2^-(K)] \non
& =& \frac{1}{2}\sum_e e \int\frac{d^3k}{(2\pi)^3}
\left[1+\frac{E_2-(\epsilon_2^e+e\mu_2)\sqrt{1+\frac{\delta M_2^4}{4\mu_2^2E_2^2}}}{\epsilon_2^e
\sqrt{1+\frac{\delta M_2^4}{4\mu_1^2E_2^2}}}\right]\coth\frac{\epsilon_2^e}{2T} \non
&& -\, \frac{\phi^2}{2}
\left[\frac{\partial \alpha}{\partial \mu_2} I_1+\frac{\partial \beta_2}{\partial \mu_2}(2I_2+J_2)\right] \, .
\eea
We have made use of the excitation energies in Eq.\ (\ref{excitekaon}) with $\delta M_1=0$. 
We have defined $I_i\equiv\frac{T}{V}\sum_K[S_{i}^+(K)+S_{i}^-(K)]$, as in Eq.\ (\ref{defIi}), and 
$J_i\equiv\frac{T}{V}\sum_K[S_{i}^+(K)-S_{i}^-(K)]$. For the explicit forms of $I_i$ and $J_i$
after performing the Matsubara sum see Eqs.\ (\ref{AppI}) and (\ref{AppJ}). Now, dropping the
vacuum contribution, adding the term from the potential $U$, and neglecting $J_2$ as well as the 
terms proportional to $\delta M_2^4$ in the square roots on the right-hand side of Eq.\ (\ref{deriv2}),
we arrive at
\begin{subequations}
\bea
 n_1 &=& \sum_e e \int\frac{d^3k}{(2\pi)^3}
f(\epsilon_1^e)-\frac{\phi^2}{2}\frac{\partial \alpha}{\partial\mu_1}I	_1  \, , \\
n_2 &\simeq& \mu_2\phi^2\left(1-
\frac{\phi^2}{4\mu_2}\frac{\partial \beta_2}{\partial\mu_2}\right) +\sum_e e \int\frac{d^3k}{(2\pi)^3}
f(\epsilon_2^e)-\frac{\phi^2}{2}\left(\frac{\partial \alpha}{\partial\mu_2}I_1+2
\frac{\partial \beta_2}{\partial\mu_2} I_2\right) \, .
\eea
\end{subequations}

\subsection{Including electric neutrality}

The electric charge density is
\be
n_Q = n_1 + 2\sum_e e \int\frac{d^3k}{(2\pi)^3} f_F(\epsilon_{\rm el}^e) \, , 
\ee
where we included the electron contribution with the Fermi distribution function 
$f_F(x)\equiv 1/(e^{x/T}+1)$ and the excitation energies 
\be
\epsilon_{\rm el}^e\equiv \sqrt{k^2+m_{\rm el}^2}-e\mu_Q \, , 
\ee
with the electron mass $m_{\rm el}$. Here, $\epsilon_{\rm el}^+$ corresponds to positrons, while 
$\epsilon_{\rm el}^-$ corresponds to electrons; in other words, $\mu_Q$ is the chemical potential for
positive electric charge, as introduced in Eq.\ (\ref{ALAR}). 
We assume that there is no lepton chemical potential. Now we can solve the self-consistency equations
(\ref{foureqs}) together with the 
neutrality condition
\be
n_Q=0 
\ee
for the variables $\phi$, $\overline{M}_1$, $\overline{M}_2$, and $\mu_Q$. In contrast to the previous 
subsections, where we have fixed $\mu_1\simeq \mu_Q + m_s^2/(2\mu)$ and $\mu_2 \simeq m_s^2/(2\mu)$, we
now solely fix the parameter $m_s^2/(2\mu)$. Then, the self-consistently determined charge chemical potential 
$\mu_Q$ determines $\mu_1$, which hence becomes a function of temperature.   
 
\begin{figure*} [ht]
\begin{center}
\hbox{\includegraphics[width=0.5\textwidth]{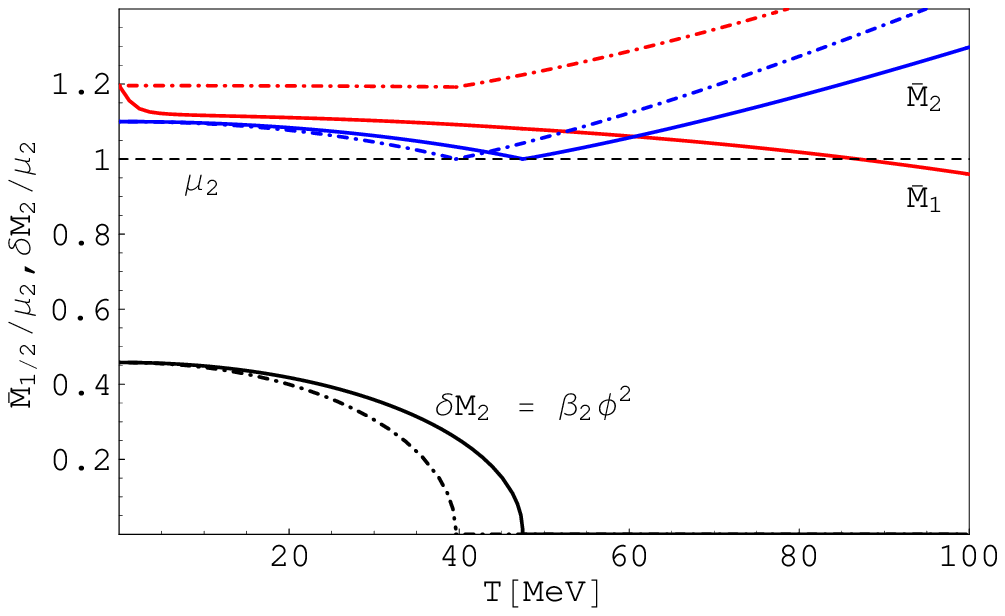}
\includegraphics[width=0.5\textwidth]{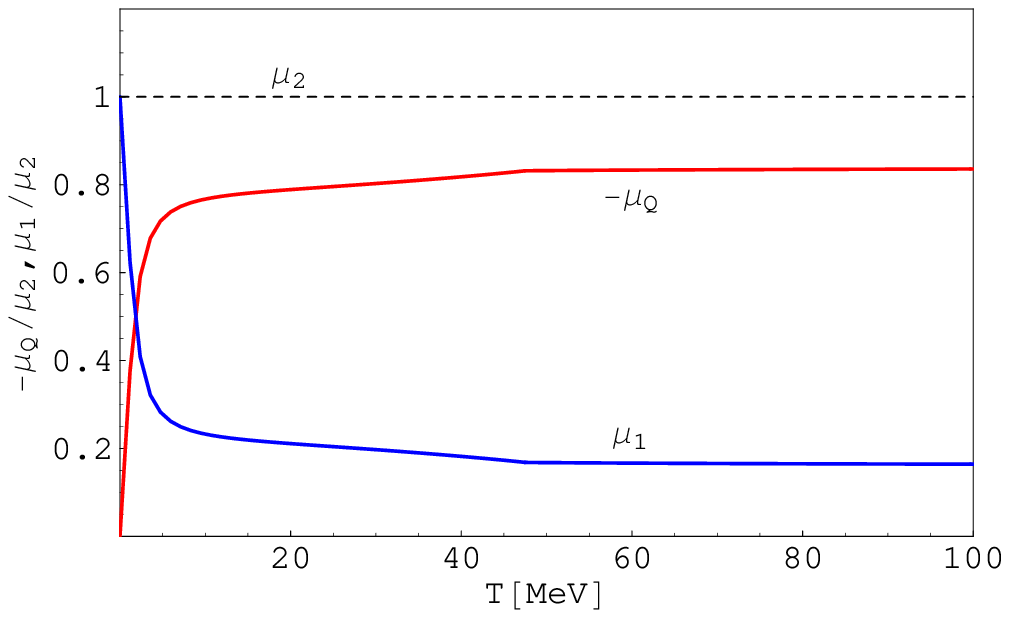}}
\caption{(Color online) 
Effect of the neutrality condition and the presence of electrons 
on the self-consistent masses and the critical temperature.
The left panel compares the masses with neutrality condition (solid) with the
ones without neutrality condition (dashed-dotted). The dashed-dotted lines are taken from 
Fig.\ \ref{figsmallmu1}. As in Fig.\ \ref{figsmallmu1}, 
for both sets of curves the bare masses are $m_1=5$ MeV, $m_2=4$ MeV, and the chemical potential of the 
condensed neutral kaon is $\mu_2=4.5$ MeV. The chemical potential of the uncondensed 
charged kaon is chosen to be identical to $\mu_2$ at $T=0$. The right panel shows the negative 
of the charge chemical 
potential $\mu_Q$, and the charged kaon chemical potential $\mu_1=\mu_Q+\mu_2$.}
\label{figneutral}
\end{center}  
\end{figure*}

In Fig.\ \ref{figneutral} we show the self-consistent masses as well as the chemical potentials 
for the neutral system. We compare this to the charged system that has the same kaon chemical potentials at 
$T=0$. We see that, compared to that system, neutrality leads to a larger critical temperature, i.e., 
the kaon condensate is more robust. One can understand this result intuitively from the
phase diagram in Fig.\ \ref{figmu1mu2T}. The phase transition lines in the upper left part of this 
phase diagram imply that the $K^0$ condensate becomes more robust for decreasing 
$\mu_{K^+}$ and fixed $\mu_{K^0}$, i.e., following a horizontal line from right to left. 
The reason for this is that the phase transition lines are not 
horizontal. In the neutral system, this is exactly how the 
chemical potentials behave as a function of temperature: 
$\mu_{K^0}$ is fixed while $\mu_{K^+}$ decreases because of an increasing 
$-\mu_Q$. Therefore, the critical temperature is larger than that of the system where both $\mu_{K^0}$ and 
$\mu_{K^+}$ are fixed (at the values that the corresponding quantities in the neutral system assume 
at zero temperature). For the given parameters, we find $T_c\simeq 47$ MeV for the neutral system, 
while $T_c\simeq 40$ MeV for the charged system.

\section{Conclusions and outlook}

We have investigated kaon condensation in the CFL phase at nonzero temperatures. This study provides
the basis for future calculations of transport properties of possible kaon-condensed CFL phases in
the interior of compact stars. We have used the effective theory which, analogous to usual 
chiral perturbation theory, contains the Goldstone modes as dynamic degrees of freedom. 

The nonzero-temperature study of this bosonic system requires a 
self-consistent scheme in order to 
compute the temperature-dependent masses. Simple approximations fail since in this case 
the melting of the condensate would lead to unphysical negative excitation energies. We have thus 
employed the 
CJT formalism, which provides self-consistent equations through stationarity conditions
for the two-particle irreducible effective action. We have treated the kaon chemical potentials as
free, but fixed, parameters and have determined the kaon masses as well as the condensate self-consistently
for arbitrary temperatures. This has been done in an approximation of the effective theory which 
reduces the Lagrangian to a form not unlike usual $\varphi^4$ theory for a scalar field. The main differences 
to $\varphi^4$ theory are the coupling constants, which turn out to be combinations of the 
kaon chemical potentials and bare masses and the presence of a second, uncondensed mode. 

Our main result is the critical temperature of condensation,
for which we have derived an analytic approximation. This high-temperature 
approximation reproduces the full numerical results to high accuracy.    
For likely values of the bare kaon masses of a few MeV and kaon chemical potentials slightly 
larger, the critical temperature is at least of the order of 20 MeV. Kaon chemical potentials of 
roughly twice the 
kaon masses lead to larger critical temperatures of approximately 80 MeV. This means that 
a (neutral) kaon condensate persists up to temperatures where fermionic excitations become important
and the effective theory is no longer valid. We conclude that, if the kaon chemical potentials 
and bare masses favor kaon condensation at $T=0$, the condensate will be present in a compact star 
which typically exhibits temperatures smaller than 1 MeV. 

Our results can be used for the calculation of several physical properties of the kaon-condensed CFL phase,
for instance the specific heat. One can also compute 
neutrino emissivities, including both condensed and thermal kaons on a consistent footing.
Both properties play an important role for the cooling properties of the
star. So far, these calculations have been done without taking into account temperature
effects in the Goldstone boson excitations \cite{Reddy:2002xc}. Furthermore, one can use our results
to compute the bulk viscosity of CFL quark matter. This has been done 
for the case where kaons do not condense
\cite{Alford:2007rw}, and for the superfluid mode alone \cite{Manuel:2007pz}.
Furthermore, one might include the effect of neutrino trapping into our calculations, 
which possibly leads to coexistence of two 
meson condensates \cite{Kaplan:2001qk}. Finally, it would be interesting to include higher orders
in the meson fields into the calculation (at zero temperature, it is easy to include {\it all} orders).

\begin{acknowledgments}
A.S.~thanks D.H.~Rischke for valuable comments. The authors acknowledge support by the 
U.S. Department of Energy
under contracts 
\#DE-FG02-91ER40628,  
\#DE-FG02-05ER41375 (OJI). 
\end{acknowledgments}

\appendix

\section{Ansatz for self-consistent boson propagator}
\label{App1}

In this appendix, we prove that the ansatz for the inverse propagator $S^{-1}$ in Eq.\ (\ref{ansatzS})
is the most general possibility. (The proof is presented for $\varphi^4$ theory, but the 
case of kaons in CFL is, within our approximations, completely analogous.) Let us derive the 
stationarity conditions without specifying any of the elements of the propagator. We denote 
the elements of the inverse propagator $S^{-1}$ as follows,
\be
S^{-1}(K) = \left(\begin{array}{cc} S_{11}^{-1}(K) & S_{12}^{-1}(K) \\ S_{21}^{-1}(K) & S_{22}^{-1}(K) 
\end{array}\right)
\qquad \Rightarrow \qquad 
S(K) = \frac{1}{{\rm det}S^{-1}(K)}\left(\begin{array}{cc} S_{22}^{-1}(K) & -S_{12}^{-1}(K) \\ 
-S_{21}^{-1}(K) & S_{11}^{-1}(K) \end{array}\right) \, .
\ee
Then, the first stationarity equation $(\ref{condmini})$ reads
\be \label{firststat}
\mu^2 = m^2 +\lambda\phi^2 +\lambda\frac{T}{V}\sum_Q\frac{3S_{22}^{-1}(Q)+S_{11}^{-1}(Q)}{{\rm det} S^{-1}(Q)}
\, ,
\ee
where we used the tree-level propagator (\ref{nondiag}).
The Dyson-Schwinger equation (\ref{schwinger}) yields for the diagonal elements
\begin{subequations}
\bea
S_{11}^{-1}(K) &=& -K^2 + m^2 + 3\lambda\phi^2-\mu^2+\lambda\frac{T}{V}\sum_Q
\frac{3S_{22}^{-1}(Q)+S_{11}^{-1}(Q)}{{\rm det} S^{-1}(Q)} \, , \label{dyson1} \\
S_{22}^{-1}(K) &=& -K^2 + m^2 + \lambda\phi^2-\mu^2+\lambda\frac{T}{V}\sum_Q
\frac{S_{22}^{-1}(Q)+3S_{11}^{-1}(Q)}{{\rm det} S^{-1}(Q)} \, , \label{dyson2}
\eea
\end{subequations}
The first of these equations, 
Eq.\ (\ref{dyson1}), can be inserted into Eq.\ (\ref{firststat}) to obtain 
$S_{11}^{-1}(K)=-K^2+2\lambda\phi^2$. This can be re-parametrized without loss of generality as 
\be \label{S11}
S_{11}^{-1}(K)=-K^2+M_+^2-\mu^2 \, .
\ee
(Yielding the condition $M_+^2-\mu^2=2\lambda\phi^2$ as one of the self-consistency equations.)
We may now subtract Eqs.\ (\ref{dyson1}) and (\ref{dyson2}) from each other,
\bea
\frac{1}{2}[S_{11}^{-1}(K)-S_{22}^{-1}(K)] &=& \lambda\phi^2-\lambda\frac{T}{V}\sum_Q
\frac{S_{11}^{-1}(Q)-S_{22}^{-1}(Q)}{{\rm det} S^{-1}(Q)} \, . \label{dyson22}
\eea
Since the self-energy is momentum independent (cf.\ its diagram in Fig.\ \ref{figv2}) the right-hand
side of this equation has to be momentum independent. Hence, 
the $K$-dependence on the left-hand side of Eq.\ (\ref{dyson22}) has to cancel out. In other words,
$S_{11}^{-1}(K)$ and $S_{22}^{-1}(K)$ must have the same momentum dependence. Choosing a 
parametrization of the momentum independent part of $S_{22}^{-1}(K)$ we thus conclude from the form 
of $S_{11}^{-1}(K)$ in Eq.\ (\ref{S11})
\be \label{S22}
S_{22}^{-1}(K)=-K^2+M_-^2-\mu^2 \, .
\ee

Having fixed the form of the diagonal elements, we turn to the off-diagonal elements of the 
Dyson-Schwinger equation. After adding and subtracting these equations to and from each other, we obtain 
\begin{subequations}
\bea
S_{12}^{-1}(K)+S_{21}^{-1}(K)&=&-2\lambda\frac{T}{V}\sum_Q
\frac{S_{12}^{-1}(Q)+S_{21}^{-1}(Q)}{{\rm det} S^{-1}(Q)} \, , \label{dyson12}\\
S_{12}^{-1}(K)-S_{21}^{-1}(K) &=& -4ik_0\mu \, .\label{dyson21}
\eea
\end{subequations}
Again, we use the fact that the right-hand side of Eq.\ (\ref{dyson12}) and thus also the 
left-hand side is momentum independent. Consequently, the momentum-dependent parts of 
$S_{12}^{-1}(K)$ and $S_{21}^{-1}(K)$ must have the same absolute value and opposite sign. 
On account of Eq.\ (\ref{dyson21})
these parts must be $\pm 2ik_0\mu$. Also on account of Eq.\ (\ref{dyson21}), 
the momentum-independent part of both off-diagonal elements must be the same. Hence, we can write
\be \label{S12S21}
S_{12}^{-1}(K) = 2ik_0\mu+c \, , \qquad S_{21}^{-1}(K) = -2ik_0\mu+c \, , 
\ee
Inserting this back into Eq.\ (\ref{dyson12}) yields 
\be
0=c\left[1+2\lambda\frac{T}{V}\sum_Q\frac{1}{{\rm det}S^{-1}(Q)}\right] \, .
\ee
The expression in square brackets does in general not vanish (note that the momentum sum vanishes for 
$T=0$).
Hence, $c=0$. 
In conclusion, Eqs.\ (\ref{S11}), (\ref{S22}), and (\ref{S12S21}) with $c=0$ define the most 
general ansatz, as used in the main part of the paper. Note that this proof is specific for
the case $\phi\neq 0$, i.e., it is valid only below the critical temperature. However, above the critical
temperature, very similar arguments apply and one obtains the same ansatz with the 
additional constraint $M_+=M_-$. 
 
\section{Matsubara sum}
\label{appB}

In this appendix we perform the Matsubara sum that is needed to derive Eqs.\ (\ref{exact}) 
for $\varphi^4$ theory.
The Matsubara sums for kaons are analogous. We rewrite the sum over bosonic 
Matsubara frequencies $q_0=-2in\pi T$ into a contour integral in the complex energy plane,
\bea
T\sum_{q_0} S^\pm(Q) &=& \frac{1}{2\pi i}\oint_C d\omega \,S^\pm(\omega)\coth\frac{\omega}{2T} \non 
&=& \frac{1}{2\pi i}\int_{-i\infty +\eta}^{i\infty +\eta} d\omega \, S^\pm(\omega)\coth\frac{\omega}{2T}
\label{contour}\, ,
\eea
where $S^\pm(\omega)$ is the propagator from Eq.\ (\ref{diagonalS}),
\be
S^\pm(\omega) = \frac{-\omega^2+q^2+M^2_\mp-\mu^2}{[\omega^2-(\e_q^+)^2][\omega^2-(\e_q^-)^2]} \, .
\ee
The closed contour $C$ encircles the poles of $\coth[\omega/(2T)]$ on the imaginary axis, such that 
no poles of $S^\pm(\omega)$ are in the enclosed region. Then, this contour can be deformed to obtain the 
second line in Eq.\ (\ref{contour}), where $\eta>0$ is infinitesimally small and 
where the symmetry of $S^\pm(\omega)$ and the antisymmetry of $\coth[\omega/(2T)]$ in $\omega$ has been used. 
Next, we can close the resulting contour in the positive half plane. This includes the positive 
poles of $S^\pm(\omega)$ and we obtain with the residue theorem
\be
T\sum_{q_0} S^\pm(Q) =\frac{1}{2}\sum_{e=\pm}\frac{1}{2\e_q^e}\left(1-e\,\frac{\overline{M}^2-M_\mp^2+2\mu^2}
{\sqrt{4\mu^2E_q^2+\delta M^4}}\right)\coth\frac{\e_k^e}{2T} \, .
\ee
Consequently,
\be \label{AppI}
T\sum_{q_0} [S^+(Q)+S^-(Q)] = \frac{1}{2}\sum_{e=\pm}\frac{1}{\e_q^e}\left(1-e\,\frac{2\mu^2}
{\sqrt{4\mu^2E_q^2+\delta M^4}}\right)\coth\frac{\e_k^e}{2T} \, ,
\ee
and 
\be \label{AppJ}
T\sum_{q_0} [S^+(Q)-S^-(Q)] = -\frac{1}{2}\frac{\delta M^2}{\sqrt{4\mu^2E_q^2+\delta M^4}}
\sum_{e=\pm}e\,\frac{1}{\e_q^e}\coth\frac{\e_k^e}{2T} \, .
\ee
With $\coth[x/(2T)] = 1+2f(x)$ and dropping the vacuum contribution one arrives at Eqs.\ (\ref{exact}) 
with the definitions (\ref{momintegrals}).

\section{Symmetry non-restoration?}
\label{AppC}

The analytical approximation of the critical temperature in Eq.\ (\ref{Tcapprox}) seems to allow
for negative expressions for $T_c^2$. In this appendix, we discuss possible implications of this 
apparently unphysical result.

The approximation (\ref{Tcapprox}) was obtained under the assumption 
$\mu_2>m_2$ (condensation of the neutral kaons). Consequently, $\alpha+2\beta_2<0$ leads to 
$T_c^2<0$. This absence of a real critical temperature 
seems to suggest that the condensate ``refuses'' to melt and persists for arbitrarily large 
temperatures. In fact, this possibility has been discussed in the literature for two-field models
which are equivalent to our expansion (\ref{Ufourthorder}) and has been termed 
``symmetry non-restoration'' (SNR), see Ref.\ \cite{SNR}.
A related phenomenon can already be seen from the original self-consistency equations
(\ref{foureqs}). Employing the high-temperature approximation (\ref{Iapprox}), we see that the self-consistent
mass $\overline{M}_1$ 
of the uncondensed mode decreases linearly in $T$ if $\alpha+2\beta_1<0$. This suggest the onset
of condensation at some large temperature, which has been termed ``inverse symmetry breaking'' 
(ISB) \cite{SNR}. Also the self-consistent mass of the condensed mode behaves unexpectedly for certain 
parameters. From Eq.\ (\ref{eq3}) we see that the temperature-dependent term proportional to $I_1\simeq I_2$
decreases linearly in $T$ if $\alpha+2\beta_2<0$. This suggests that, for large temperatures, 
the condensate $\phi$ has to {\em increase} with increasing $T$ in order to avoid a negative mass squared. 
On the other hand, we know that our approximation breaks down for large condensates because we have
employed an expansion in $\phi$, see comments below Eq.\ (\ref{Ufourthorder}). 
Therefore, in our specific context, the exotic possibilities of SNR and/or ISB might be an artifact of the 
approximation, and we leave it to future studies to decide whether they remain a viable option in the
full effective theory.

\begin{figure} [ht]
\begin{center}
\includegraphics[width=0.5\textwidth]{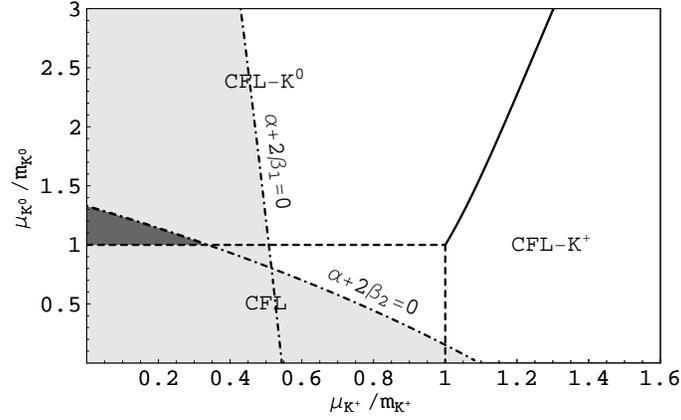}
\caption{
Parameter regions where symmetry non-restoration (dark grey) and/or inverse symmetry 
breaking (light grey) is suggested from the approximate effective theory. The area shaded in dark grey 
only appears for sufficiently large differences in the kaon masses. For this plot we have chosen $m_1=5m_2$. 
} 
\label{figISB}
\end{center}
\end{figure}

Instead, we simply identify the parameter regions where $\alpha+2\beta_i<0$. 
In Fig.\ \ref{figISB} we plot the 
(dashed-dotted) lines $\alpha+2\beta_1=0$ and $\alpha+2\beta_2=0$ in the 
$\mu_1$-$\mu_2$-phase diagram. These lines have 
been obtained with the help of the definition (\ref{alphabeta}). The shaded
areas to the left of (and below) these curves correspond to $\alpha+2\beta_i<0$. 

Regions shaded in light
grey indicate ISB, i.e., a decreasing self-consistent mass of an uncondensed mode, suggesting 
high-temperature condensation for this mode when the mass is sufficiently small. This does not occur for 
the parameters chosen in Fig.\ \ref{figsmallmu1}, whereas the phase diagram in 
Fig.\ \ref{figmu1mu2T} contains such 
parameter regions. However, the smallness of the coupling constants $\alpha$ and $\beta_{i}$ 
prevents the self-consistent mass from becoming close to zero. Hence, a possible onset of condensation
would appear at temperatures far beyond the validity of the effective theory, rendering the question 
of ISB for the given parameters purely theoretical. 

The region shaded in dark grey corresponds to a negative critical temperature squared, i.e., possibly SNR.  
The occurrence of this region requires
a sufficiently large difference in the bare kaon masses $m_1^2>12m_2^2$. The line $\alpha+2\beta_2=0$ 
intersects the line $\mu_2=m_2$ at the point $(\mu_1,\mu_2)=(\sqrt{m_1^2-3m_2^2}-3m_2,m_2)$. 
Thus, this 
region is not present in the phase diagram in Fig.\ \ref{figmu1mu2T}, where the kaon masses 
are chosen to be sufficiently close together.
Because of large uncertainties in these masses a large difference in kaon 
masses is not excluded. Hence we cannot exclude the possibility 
of exotic phenomena, and future studies going beyond our simple expansion in the fields are 
required.

\end{document}